\documentclass[useAMS,usenatbib]{mn2e}

\usepackage{psfig}
\usepackage{amssymb}
\usepackage{multirow}
\usepackage{color}

\newcommand\halpha{H$\alpha$}

\newcommand\Halpha{H$\alpha$}
\newcommand\Hbeta{H$\beta$}

\newcommand\kms{km s$^{-1}$}

\title[SDSS galaxies mapping]
{Beyond the fibre: Resolved properties of SDSS galaxies
\thanks{Based on observations made with ESO Telescopes at the Paranal
  Observatory under programmes 076.B-0408(A) and 078.B-0194(A).} }

\author[J.Gerssen]{J.~Gerssen$^{1}$\thanks{E-mail:
jgerssen@aip.de}, D.~J.~Wilman$^{2}$, L.~Christensen$^{3}$\\
$^{1}$Leibniz-Institut f\"ur Astrophysik Potsdam, An der Sternwarte 16, 14482 Potsdam, Germany.\\
$^{2}$Max-Planck-Institut f\"ur Extraterrestrische Physik, Giessenbachstraße, 85748 Garching, Germany.\\
$^{3}$Excellence Cluster Universe, Technische Universit\"at M\"unchen, Bolzmanstrasse 2, 85748 Garching, Germany\\
}

\begin{document}

\maketitle

\label{firstpage}

\begin{abstract}
  
We have used the VIMOS integral field spectrograph to map the emission
line properties in a sample of 24 star forming galaxies selected from
the SDSS database. In this data paper we present and describe the
sample, and explore some basic properties of SDSS galaxies with resolved
emission line fields. We fit the \Halpha+[NII] emission lines in each
spectrum to derive maps of continuum, \Halpha\ flux, velocity and
velocity dispersion. The \Halpha, \Hbeta, [NII] and [OIII] emission
lines are also fit in summed spectra for circular annuli of increasing
radius. A simple mass model is used to estimate dynamical mass within 10
kpc, which compared to estimates of stellar mass shows that between 10
and 100\% of total mass is in stars.  We present plots showing the
radial behaviour of EW[\Halpha], u-i colour and emission line ratios.
Although EW[\Halpha] and u-i colour trace current or recent star
formation, the radial profiles are often quite different.
Whilst line ratios do vary with annular radius, radial gradients in
galaxies with central line ratios typical of AGN or LINERS are mild,
with a hard component of ionization required out to large radii. We use
our VIMOS maps to quantify the fraction of \Halpha\ emission contained
within the SDSS fibre, taking the ratio of total \Halpha\ flux to that
of a simulated SDSS fibre. A comparison of the flux ratios to
colour-based SDSS extrapolations shows a 175\% dispersion in the
ratio of estimated to actual corrections in normal star forming
galaxies, with larger errors in galaxies containing AGN. We find a
strong correlation between indicators of nuclear activity: galaxies with
AGN-like line ratios and/or radio emission frequently show enhanced
dispersion peaks in their cores, requiring non-thermal sources of
heating.
Altogether, about half of the galaxies in our sample show no evidence
for nuclear activity or non-thermal heating.
The fully reduced data cubes and the maps with the line fits results are
available as FITS files from the authors.

\end{abstract}


\section{Introduction}

The unprecedented large number of galaxies observed in the Sloan
Digital Sky Survey (SDSS, York et al. 2000) has made a significant
impact on our quantitative and qualitative understanding of galaxy
formation and evolution.  The relationships between the derived
physical parameters from the SDSS galaxies, such as the star formation
rates and metallicities, describe how efficiently galaxies turn their
gas into stars and help constrain theoretical modelling of star
formation and galaxy evolution.  Unfortunately, as SDSS spectra are
fibre-based, these astrophysical parameters are centrally averaged
quantities only (Kewley et al. 2001, 2005; Brinchmann et al. 2004
(hereon B04); Wilman et al. 2005).

To avoid characterization of the global galaxy properties based on
{\it extrapolated} central measurements requires spatially resolved
observations.  Integral Field Spectroscopy (IFS) is an excellent tool
to map the internal structure of galaxies and to further our knowledge
of these systems in the local Universe.  Surveys of nearby galaxies
with IFS include SAURON (Bacon et al. 2001) to study early type
galaxies, PINGS (Rosales-Ortega et al. 2010) to map the properties of
disk galaxies and CALIFA (S\'anchez et al.  2011) an ambitious IFS
survey of some 600 galaxies spanning all Hubble types in the local
Universe.

A related project, GHASP (Garrido et al. 2002; Epinat et al. 2010),
maps the \Halpha\ velocities in spiral galaxies using the scanning
Fabry-Perot technique in a sample of 153 nearby spiral galaxies and
build up a local reference sample for comparison with high-z
observations.

The SDSS database is the largest source of extra-galactic properties,
albeit centrally averaged quantities only.  It has a fairly uniform
coverage of galaxies in the redshift range out 0.1.  However, most IFS
surveys cover a much more local volume, e.g. CALIFA goes out to $z <
0.03$.
In this paper we present an IFS study to map the properties in a
sample of 24 galaxies selected from the SDSS database.  These systems
have EW[\halpha]$>20$\AA\ and are uniformly distributed over the
redshift range 0 to 0.1.  This allows us to systematically compare the
centrally averaged values to the integrated properties, address
aperture bias in star forming galaxies, and characterize the large
scale structure of these systems in detail.

The layout of this paper is as follows.  In section 2 we describe the
criteria used to construct our sample of 24 SDSS galaxies.  In
section 3 we describe the data reduction process and the emission line
analysis. We present the radial gradients and 2D maps of \halpha\
derived from these data and interpret them in section 4 and 5 respectively.

\section{Sample}

\begin{figure*}
 \centerline{\psfig{figure=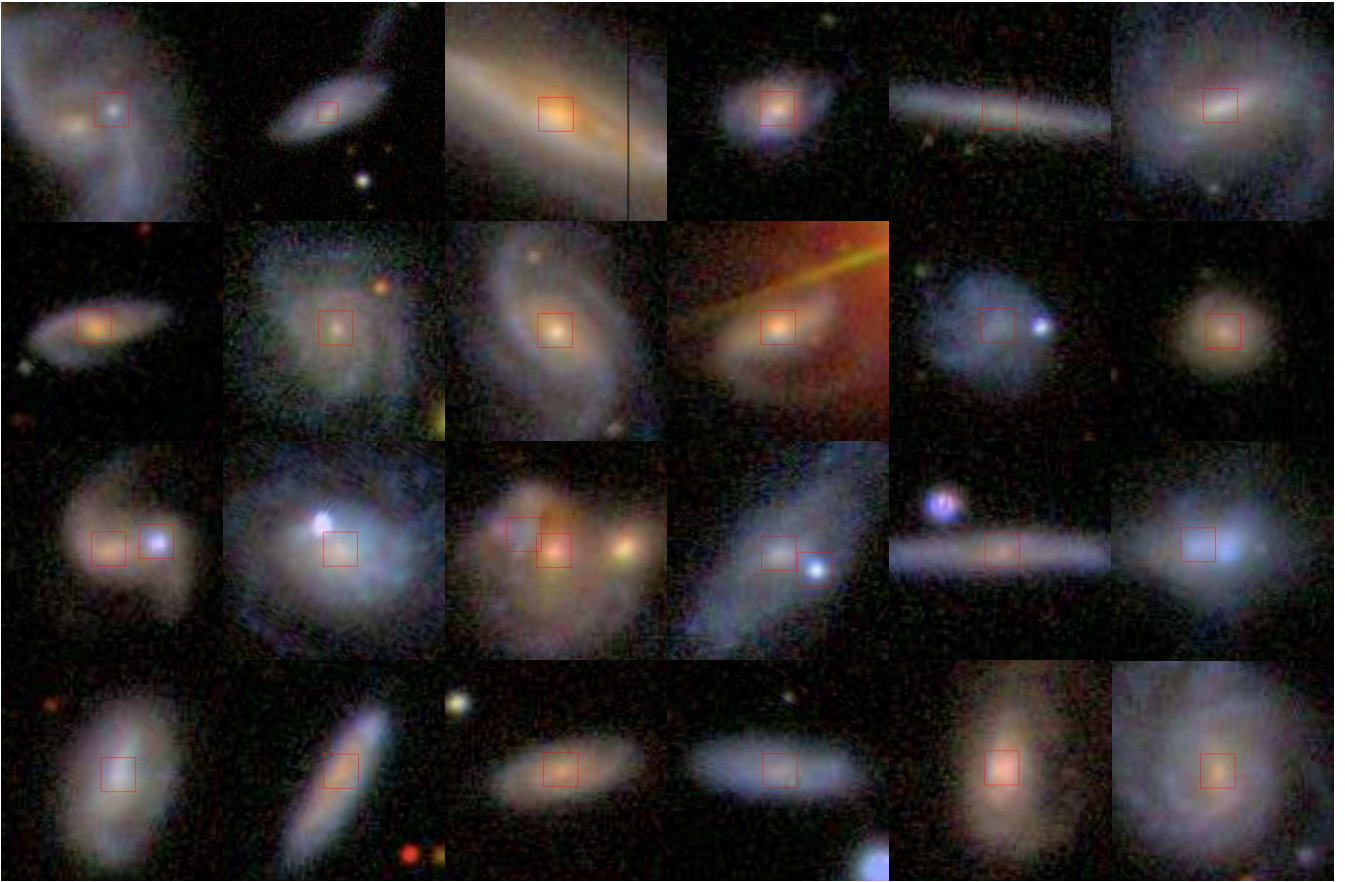,width=\textwidth}}
 \caption{Colour SDSS imaging of our selected galaxies. Top to bottom, left to right,
   galaxies 01-06, 07-12, 13-18, 19-24. Red squares show the location of
   SDSS spectral fibres.}
\label{f:sdssim}
\end{figure*}

Targets were selected from the SDSS Data Release 4
(DR4).  The only criteria for selection were
EW[H$\alpha$]$\geq20$\AA; model r-band magnitude, $r \leq 17$;
avoidance of objects with sizes much smaller or larger than the FOV
(27\arcsec$\times$27\arcsec, but this covers a range of physical size
at variable redshift) and suitable right ascension and declination for
observation.
We selected a total of 24 galaxies.  For visual reference purposes we
present a mosaic of the SDSS colour images in Fig.~\ref{f:sdssim}.

\begin{figure*}
 \centerline{\psfig{figure=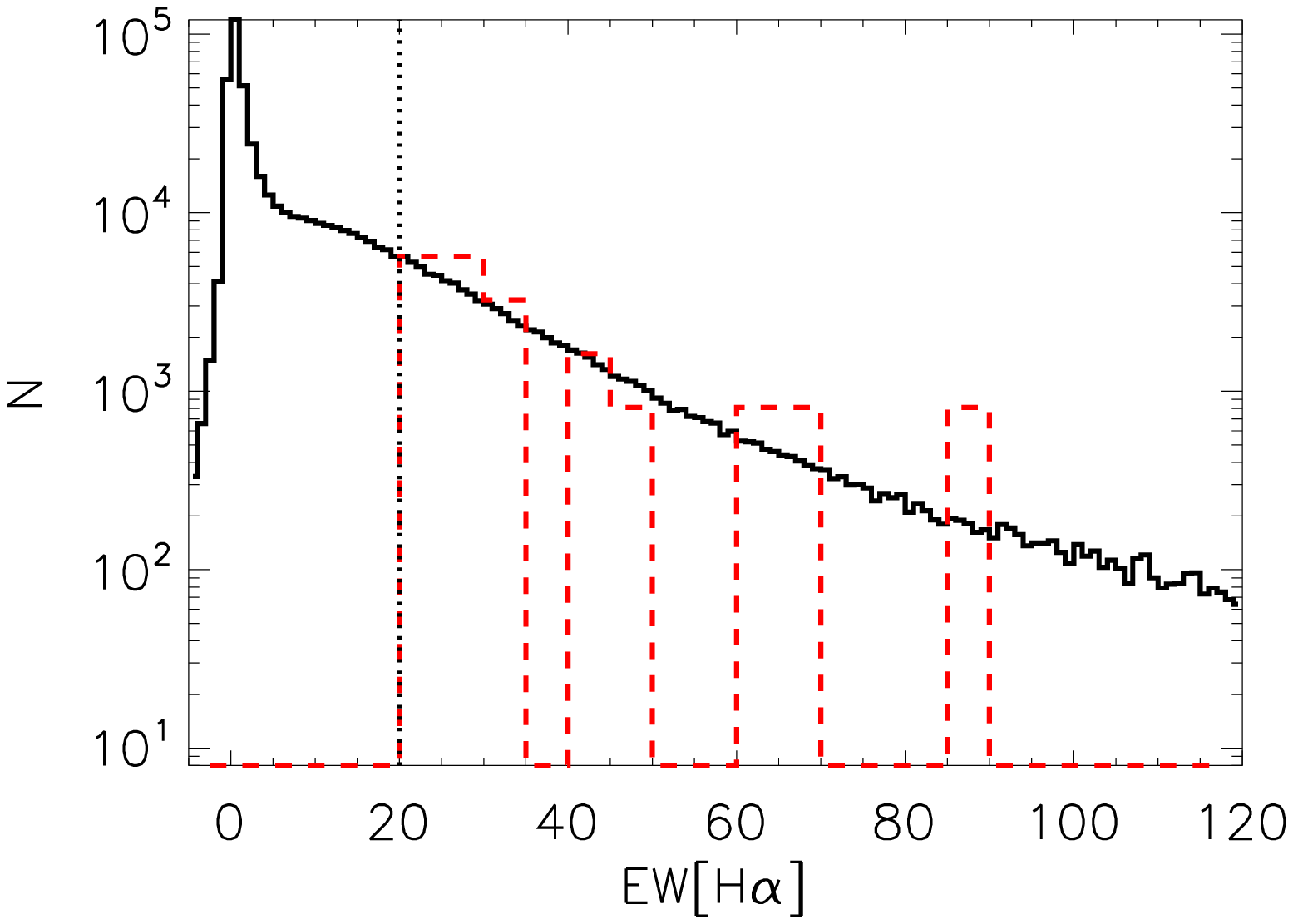,width=0.45\textwidth}\psfig{figure=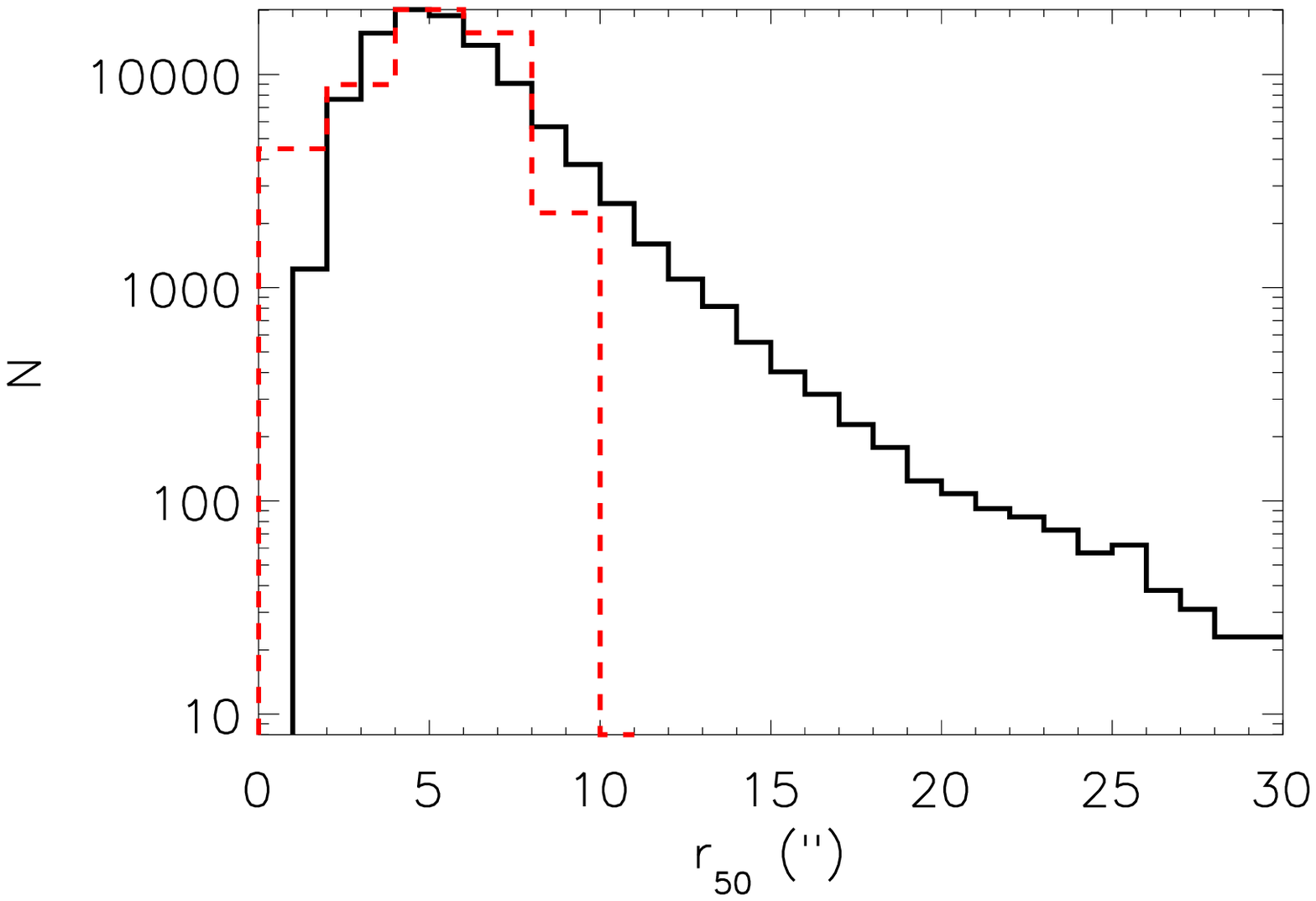,width=0.45\textwidth}}
 \centerline{\psfig{figure=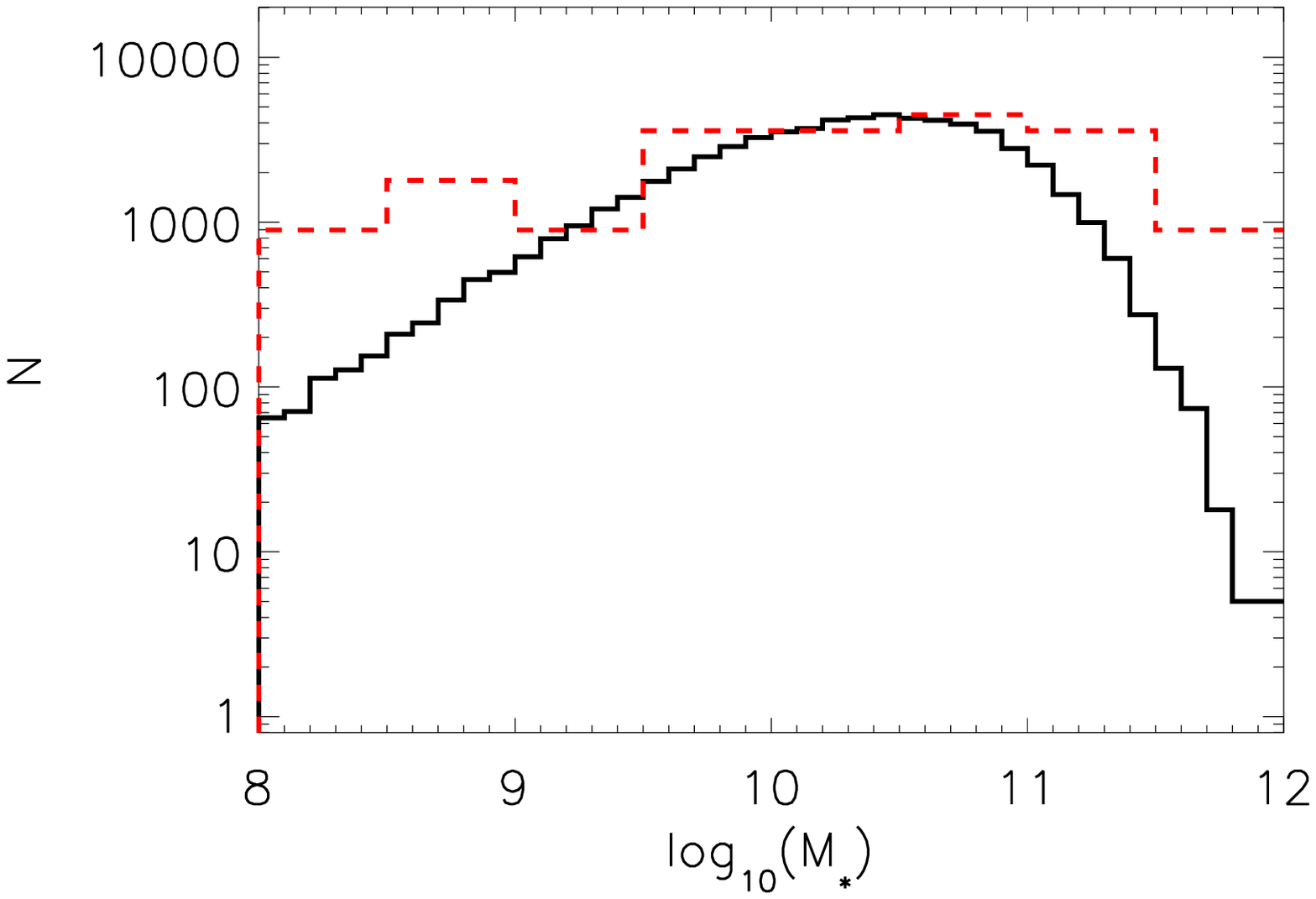,width=0.45\textwidth}\psfig{figure=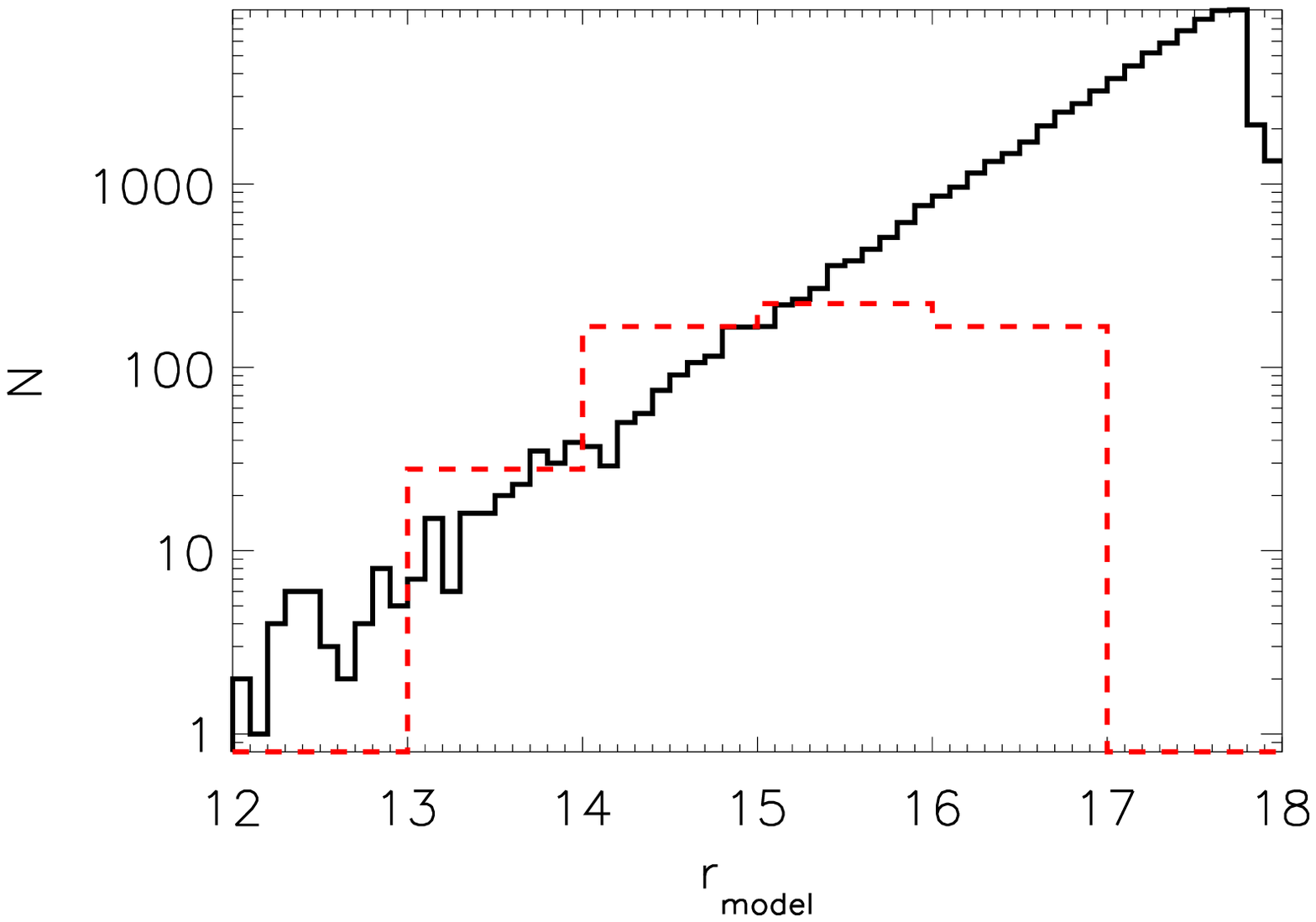,width=0.45\textwidth}}
 \centerline{\psfig{figure=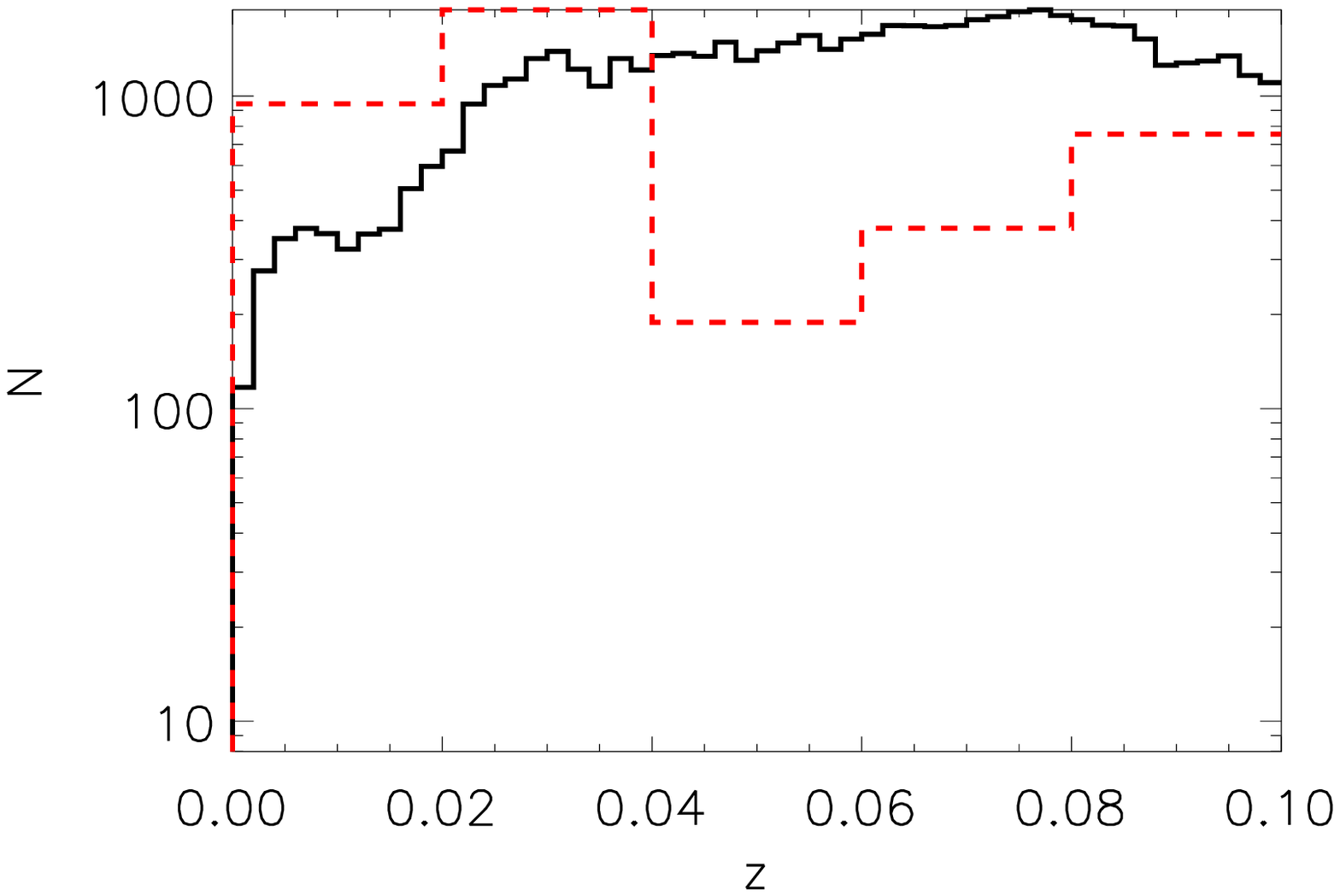,width=0.45\textwidth}}
 \caption{Top left: Distribution of EW[H$\alpha$] (B04, fibre
     measurements) for all SDSS DR4 galaxies (black, solid
   histogram) and our VIMOS targets (red, dashed histogram,
   scaled). By selection, we are sampling the more highly star forming
   galaxies (EW[H$\alpha$]$=20$\AA, dotted vertical line).  Top right:
   Distribution of $r_{50}$, the r-band Petrosian half-light radius in
   arcseconds. The solid black histogram represents all galaxies with
   EW[H$\alpha$]$\geq20$\AA, whilst our targets are represented by the
   dashed red histogram (renormalized). Likewise, the middle and
   bottom panels illustrate the derived stellar mass \citep{gal06},
   r-band model magnitude and redshift distributions of the
   EW[H$\alpha$]$\geq20$\AA\ population and our targets respectively.}
\label{f:selection}
\end{figure*}

As demonstrated in the top left panel of Fig.~\ref{f:selection}, the
primary selection criterion of EW[H$\alpha$]$\geq20$\AA\ selects some
of the more strongly H$\alpha$ emitting galaxies.
\footnote{EW[H$\alpha$] is taken from B04 fibre measurements.}  This
corresponds to the upper $20\%$ of all galaxies in EW[H$\alpha$].  The
peak at EW[H$\alpha$]$\sim 0$\AA\ can be fit using a Gaussian whose
width should represent the typical error on this measurement for a
passive galaxy. Subtracting this population leaves only star-forming
galaxies, of which our EW[H$\alpha$]$\geq20$\AA\ criterion selects the
upper $27\%$.

It is also important to note that
this selection is applied in terms of the {\it fibre} or {\it central}
EW[H$\alpha$], and so it is possible that we miss objects with most of
the line flux in the outer regions, such as face-on spiral galaxies
with a passive bulge and highly star forming outer disk.

The other four panels of figure~\ref{f:selection} show how our targets
(red, dashed histograms) are distributed in the radius containing 50\%
of the Petrosian r-band flux ($r^{''}_{50}$), stellar mass ($M_*$,
Gallazzi et al, 2006), r-band model
magnitude and redshift compared to the overall SDSS DR4 distribution of
EW[H$\alpha$]$\geq20$\AA\ objects (black, solid histograms). We cover
this parameter space, excluding only the galaxies with large apparent
sizes ($r_{50} >10\arcsec$) for which we would not have been able to
observe close to the whole galaxy in one VIMOS pointing.

\begin{table*}
\centering
\begin{minipage}{140mm}
\caption{Properties of the galaxies in our VIMOS IFU sample derived from SDSS DR5. 
Listed in column six are luminosity distances. The SDSS derived $r$-band model magnitude
and Petrosian radius are listed in columns eight and nine respectively.}
\begin{tabular}{@{}llcccccccc@{}}
 \hline
 Name & SDSS name & RA & Dec & $z$  &$D_L^{1)}$ & scale & $r_{\rm model}$ & $r_P$ & EW[\Halpha]  \\ 
 &     & (J2000) & (J2000) & & Mpc & kpc/$\prime\prime$ & & (arcsec) & \\ 
 \hline
% p76 sample
 gal01 & SDSS J081753+002235 & 08h17m53s  & +03d22m35s & 0.030     & 129.7 & 0.593 & 15.52 & 16.43 & 20.08 \\
 gal02 & SDSS J085555+034530 & 08h55m55s  & +03d45m30s & 0.028     & 120.8 & 0.554 & 15.50 & 10.43 & 34.33 \\
 gal03 & SDSS J091750$-$001642 & 09h17m50s  & $-$00d16m42s & 0.017 & 72.7  & 0.341 & 13.87 & 20.69 & 21.72 \\ 
 gal04 & SDSS J091844+073713 & 09h18m44s  & +07d37m13s & 0.106     & 484.1 & 1.919 & 16.16 &  5.88 & 30.05 \\
 gal05 & SDSS J092001+070322 & 09h20m01s  & +07d03m22s & 0.012     &  51.2 & 0.242 & 16.17 & 14.06 & 29.26 \\
 gal06 & SDSS J092348+020645 & 09h23m48s  & +02d06m45s & 0.024     & 103.3 & 0.477 & 14.68 & 17.98 & 25.68 \\
 gal07 & SDSS J092950+045021 & 09h29m50s  & +04d50m21s & 0.097     & 440.3 & 1.774 & 16.20 &  8.09 & 20.19 \\
 gal08 & SDSS J093148+072405 & 09h31m48s  & +07d24m05s & 0.034     & 147.4 & 0.668 & 15.26 & 14.98 & 21.94 \\
 gal09 & SDSS J095411+070738 & 09h54m11s  & +07d07m38s & 0.041     & 178.7 & 0.799 & 14.74 & 13.77 & 24.23 \\
 gal10 & SDSS J095432+064902 & 09h54m32s  & +06d49m02s & 0.074     & 330.5 & 1.389 & 15.43 & 10.06 & 31.35 \\
 gal11 & SDSS J095728+074044 & 09h57m28s  & +07d40m44s & 0.022     &  94.5 & 0.439 & 16.30 & 10.62 & 30.73 \\
 gal12 & SDSS J095732+055309 & 09h57m32s  & +05d53m09s & 0.093     & 420.9 & 1.708 & 16.52 &  6.96 & 22.35 \\
% p78 sample
 gal13 & SDSS J095344$-$000526 & 09h53m44.3s & -00d05m26s & 0.084  & 377.8 & 1.559 & 15.79 & 10.32 & 35.44 \\
 gal14 & SDSS J090656$-$000153 & 09h06m55.5s & -00d01m53s & 0.019  &  81.4 & 0.380 & 14.94 &  8.88 & 44.52 \\
 gal15 & SDSS J095940+003512 & 09h59m39.5s & +00d35m12s & 0.066    & 293.0 & 1.250 & 15.07 & 13.58 & 25.47 \\
 gal16 & SDSS J090145+003119 & 09h01m45.2s & +00d31m19s & 0.019    &  81.4 & 0.380 & 15.41 & 16.94 & 46.25 \\
 gal17 & SDSS J091844+012341 & 09h18m43.6s & +01d23m41s & 0.037    & 160.8 & 0.725 & 15.69 & 14.78 & 25.84 \\
 gal18 & SDSS J091304+020219 & 09h13m04.2s & +02d02m19s & 0.013    &  55.5 & 0.262 & 15.22 & 11.55 & 67.85 \\
 gal19 & SDSS J095317+015841 & 09h53m17.0s & +01d58m41s & 0.021    &  90.1 & 0.419 & 15.09 & 10.24 & 26.92 \\
 gal20 & SDSS J095523+023459 & 09h55m23.4s & +02d34m59s & 0.031    & 134.1 & 0.612 & 15.40 & 10.67 & 41.93 \\
 gal21 & SDSS J095727+003159 & 09h57m27.0s & +00d31m59s & 0.082    & 368.3 & 1.525 & 15.90 &  8.99 & 23.52 \\
 gal22 & SDSS J092407+020457 & 09h24m06.5s & +02d04m57s & 0.024    & 103.3 & 0.477 & 15.88 & 10.34 & 24.98 \\
 gal23 & SDSS J095228+014121 & 09h52m27.8s & +01d41m21s & 0.074    & 330.5 & 1.389 & 14.69 &  9.79 & 59.73 \\
 gal24 & SDSS J095414+003818 & 09h54m14.1s & +00d38m18s & 0.035    & 151.9 & 0.687 & 14.64 & 16.54 & 28.75 \\
 \hline
\multicolumn{7}{l}{1) Calculated assuming: $\Omega_\Lambda$=$0.73$, $\Omega_M$=$0.27$, $H_0$=$71$ km s$^{-1}$ Mpc$^{-1}$.} \\
 \label{t:sample}
\end{tabular}
\end{minipage}
\end{table*}

\section{Reduction \& Analysis}
\label{s:reduction}

The observations were obtained with the VIMOS spectrograph mounted on
the VLT-{\it Melipal} (programme IDs 076.B-0408(A) and 078.B-0194(A))
in IFU mode. Each of the 40$\times$40 spaxels (spatial element pixels)
of the VIMOS IFU samples 0\farcs67 on the sky. The 1600 spectra are
separated into four channels of VIMOS and recorded on four
2k$\times$4k pixel EEV CCDs.  The spectral resolution obtained with
the medium resolution (MR) grism is $R\sim$720, and the wavelength
coverage is 4800--9300 {\AA}. For each of the galaxies, two exposures
of 2000~s (gal01-12) or 2040~s (gal13-24) were obtained with
offsets of a few arcseconds between the pointings. Calibration data
needed for the data reduction were obtained following the science
exposures.

The data were reduced using IDL scripts adapted for VIMOS data (Becker
2002; S\'anchez et al. 2005), as well as custom-made IDL scripts, which
handled separately the four files from the four quadrants. After bias
subtraction, the locations of the spectra on the detectors were
determined from calibration data obtained with a continuum lamp
exposure.  Each quadrant and exposure was checked interactively for
shifts between the science frame and the traces of each fibre derived
from the flat field frames taken immediately before and after the
science exposure. In a few cases, the traces determined by the software
were corrected interactively.

The continuum, an arc line exposure, and the science data were extracted
after cosmic ray removal (Pych 2004). The wavelength solution was
determined for each of the 1600 spectra using standard IRAF routines. To
correct for wavelength shifts from one spaxel to the next, the strong
5577 \AA\ sky emission line was fit by a Gaussian function to find the
wavelength of the peak. Any detected shift was applied to the wavelength
solution and the resulting accuracy was within 10\% of a pixel.

To facilitate the subsequent analysis we rebinned individual spectra
onto a common wavelength scale: a wavelength range from 4800 to 9300
\AA\ with a dispersion per pixel of 2.53 \AA.

To flat-field the data, we used the extracted continuum lamp exposure
to create a normalised relative transmission correction by dividing
each of the 1600 spectra by the average continuum spectrum. Any
remaining spaxel-to-spaxel transmission variations were corrected using
the 5577 \AA\ sky emission line flux in each spaxel. The derived
best-fit Gaussian line fluxes were used to create a normalized
throughput map that was used as a correction to the spatial flat field.
The extracted, wavelength calibrated data were divided by this
normalised flat field. Finally, the spectra were collected into a
three-dimensional data cube.

Sky subtraction was performed by visually identifying those spaxels
dominated by sky background in each data cube.  A sky spectrum was
constructed from each exposure by taking the median spectrum over the
selected sky spaxels which was then subtracted from all spectra in the
data cube.

To calibrate flux as a function of wavelength we use the flux calibrated
SDSS spectra of our target objects themselves.  Simulated SDSS spectra
are created from our VIMOS data cubes using a `software SDSS
fibre'. That is, we create a circular aperture with a radius of
1\farcs5, centered on the position of the SDSS fibre on the sky and
convolved to the seeing at which the SDSS spectrum is observed.  The
VIMOS spectra within this aperture are summed to create a simulated SDSS
spectrum.  The SDSS spectra are rebinned to the wavelength scale of our
VIMOS data.  The flux correction as a function of wavelength is the
ratio of the simulated SDSS spectrum to the actual, flux-calibrated SDSS
spectrum (see Fig.~\ref{f:reduction}).  We linearly interpolate over
emission line regions.  Our resultant flux-calibrated VIMOS spectra are
therefore calibrated directly to the {\it continuum} of the SDSS
spectra. This helps to avoid systematic differences when comparing {\it
  line} flux measurements in section~\ref{s:apeffects}.

Some of the observations were obtained at relatively high airmass, and
clearly show the effects of atmospheric differential refraction over the
long wavelength range. It is necessary to correct for this effect before
the data cubes are combined. First, the location of the centre of the
galaxy as a function of wavelength was found by cross correlating an
image slice in the data cube around 5000~{\AA} with consecutive slices
(or wavelength channels) of 10~{\AA} width.  Polynomial functions were
fit versus the x- and y coordinates. Then the data were resampled by
shifting each two dimensional image in the data cube by a pixel fraction
indicated by the polynomial fits. This resampling used drizzling scripts
\citep{fruchter02}. Spatial offsets between the two exposures were
determined and the two cubes were combined by averaging the data into a
single data cube for further analysis.

Due to the change in dispersion from fibre to fibre combined with strong
detector fringes in the red, sky lines are not all well subtracted, and
significant residuals are present redwards of 7230 {\AA}. All emission
lines analysed in this paper, apart from H$\alpha$ and [NII] from gal04,
lie bluewards of this. Because the emission lines are much brighter than
the continuum sky background, sky subtraction errors are not significant
for the further analysis of the emission lines. In the continuum images
constructed from the full data cube, the sky line residuals produce
additional noise patterns when galaxy emission is similar to the
background, as can be seen e.g. at the edges of some galaxies in the
left-most column in Fig. 9.

\subsection{Continuum subtraction}
\label{s:contsub}

The continuum levels in these data are faint as can be seen in the
middle panel of Fig.~\ref{f:reduction}.  As our sample is selected to
have relatively large EW ($\ge 20$ \AA) in \Halpha\ emission this faint
underlying absorption does not measurably affect the best-fit emission
line parameters.

However, we also analyse spectra in radially binned annuli (sections 4.1
\& 4.2 and Table~\ref{t:flux}).  In this case the increase in S/N ratio
in these stacked spectra allows us to fit their stellar continuum using
the stellar absorption line fitting software {\tt Paradise} (Walcher et
al. 2009).  Briefly, this code uses a set of Bruzual \& Charlot (2003)
spectra as templates and finds their combination that best matches the
input spectrum in a least square sense.  (The code is based on the
method of Rix and White, 1992.) Spectral regions dominated by emission
lines or artefacts are flagged and are not used in the fit.  We subtract
the best-fit template from the input spectrum to correct for underlying
absorption.  The code also creates a smoothed version (using a running
mean) of the continuum level in the input spectrum to facilitate
equivalent width analysis.

We have analysed the \Hbeta\ emission line flux in original and in the
continuum subtracted spectra separately.  On average, over all annuli
and galaxies, the EW-\Hbeta\ derived in the continuum subtracted spectra
is larger by $2.3$ \AA\ (with a 1-sigma scatter of $1.5$\AA).  This is
consistent with the oft adopted EW$_{\rm abs}$=2 \AA\ correction value
for underlying absorption (e.g. Mattsson \& Bergvall, 2009).  In the
non-binned spectra (section~\ref{s:maps}) we do not attempt to fit the
underlying absorption but instead apply an EW=2\AA\ correction directly.
This correction is relevant only to line-flux maps presented in
section~\ref{s:fluxmaps}.

\subsection{Emission line fitting}

We analyse the emission lines in all data cubes using a custom built
fitting tool. This tool is written in IDL and makes use of Craig
Markwardt's mpcurvefit.pro fitting
routine\footnote{http://www.physics.wisc.edu/\textasciitilde
craigm/idl/fitting.html}.  Our line fitting tool fits Gaussian
components to $n$ lines simultaneously and can be run interactively or
automatically over all spectra in a data cube. The interactive option
allows us to both inspect the best-fit results visually and to re-fit
emission lines with starting parameters adjusted and/or kept fixed.

In practice we have set $n=3$ to fit the \halpha+[NII] lines
simultaneously.  We make the assumption that these emission lines trace
the same underlying gas kinematics. We therefore
tie their centroids and widths to a common value. That is, we fit one
centroid and one line width per set of $n=3$ lines. Only the line flux
is a free parameter for each individual line. We simultaneously fit the
underlying continuum using a constant value. The free parameters in
each fit are: the centroid, dispersion (line width), continuum level,
and $n$ $\times$ line flux.  Where possible we fit the \Hbeta+[OIII]
lines with a similar procedure.

Reducing each spaxel/spectrum from the counts distributed on the
detector to a fully reduced 1D spectrum relies on several interpolation
steps.  Propagating errors through more than one interpolation steps is
in practice not tractable.  The optimal data reduction would thus only
require a single interpolation step from raw data on the detector to a
fully reduced data cube (e.g. Weilbacher et al. 2009).

As such an approach is not feasible with VIMOS data we estimate the
errors using a two step process in this paper.  First we apply our
Gaussian fitting procedure to raw spectra. That is, spectra that
have been extracted from the detector and collapsed to 1D spectra (a
step that typically requires interpolation) but have not been subjected
to other interpolation steps.  In these spectra count statistics are,
to first order, well defined.  Hence, when we use the count values to
weigh the input spectrum the resultant best-fit errors are a proper
estimate of the one-sigma uncertainty on the best-fit values.  This
allows us to define the relative error ($\delta_i$) for every spaxel
with sufficient flux.  When, in the second step, we apply the line
fitting procedure to the fully reduced spectra we can now use the
relative error derived for that spectrum from the corresponding raw
spectrum to estimate the error, e.g.  $\Delta$Flux$_i$ = Flux$_i$
$\times \ \delta_i$.

\begin{figure*}
 \centerline{\psfig{figure=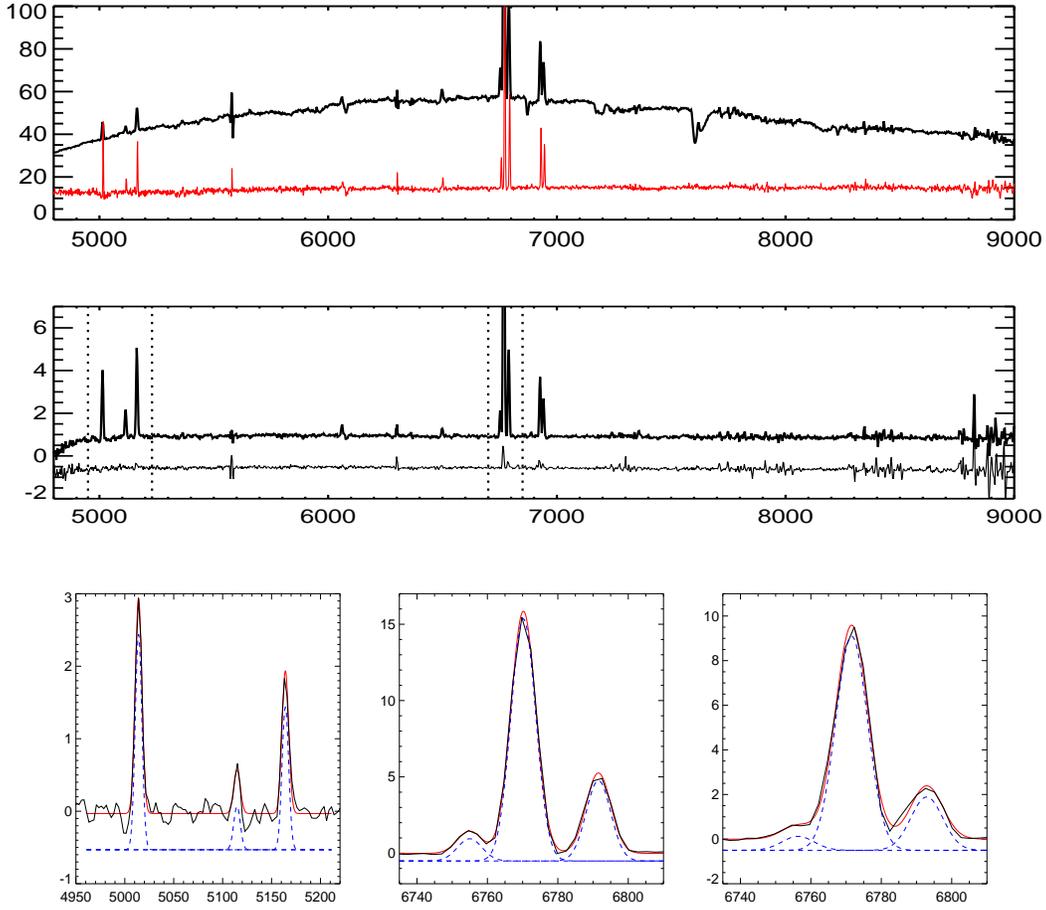,width=0.85\textwidth}}
 \caption{ Examples showing intermediate results, reduced spectra, and
   detailed views of the \Hbeta\ and \Halpha\ emission line regions.  In
   all panels the horizontal axes are in Angstroms and the vertical axes
   show flux in units of 10E-17 erg/s/cm2/\AA.  {\bf Top panel:} The
   black line shows the spectrum derived for a simulated SDSS aperture
   in an intermediate reduced data cube (gal20 is shown in this example
   -- arbitrary flux units).  The flux calibrated SDSS spectrum itself
   is shown in red.  By matching the simulated spectrum to the SDSS
   spectrum wavelength by wavelength, see section~\ref{s:reduction}, we
   can remove both the VIMOS response profile and establish the flux
   calibration scale.  {\bf Middle panel:} Reduced spectra along two
   individual spaxels in gal20. The upper line shows an example of a
   bright spectrum along a central spaxel while the bottom line shows a
   faint spectrum in a spaxel near the edge of the galaxy.  Note that
   the faint spectrum has been vertically offset by -1.0 for clarity.
   In the subsequent emission line analysis we only use the two
   wavelength regions bounded by the dotted lines and is thus not
   affected by the sky subtraction residuals seen beyond 7000 \AA.
   {\bf Bottom panels:} Close-ups views of emission line fits in
   continuum subtracted gal20 spectra.  The left and middle panels show
   the \Hbeta+[OIII] and \Halpha+[NII] complex respectively for a
   spectrum binned in a circular annulus with a radius of $2''$. The
   best-fit is oveplotted in red and the individual best-fit components
   are shown in blue (with a negative offset for clarity).  The
   centroids and the width of three individual Gaussian components in
   each panel are tied to each other.  That is, we assume that the gas
   traces the same kinematic features.  Hence, we fit only one systemic
   velocity and dispersion per panel.
   For comparison the right most panel shows the \Halpha+[NII] in the
   annulus at the largest radius ($10''$).  Averaging over lines with
   up to 400 km/s velocity difference at this radius leads to a
   noticeable broadening but the overall shape remains Gaussian.  }
\label{f:reduction}
\end{figure*}

The line fitting method is illustrated in the bottom panels of
Fig.~\ref{f:reduction}.  Close ups of the \Hbeta+[OIII] and
\Halpha+[NII] lines are shown with the best-fit model overplotted.
Shown here are gal20 spectra that have been derived in the smallest
radius annulus and are continuum subtracted.
For comparison, we also show the \Halpha+[NII] complex in the largest
radius annulus.  In this case we average over emission lines with up to
400 km/s velocity difference. While this results in a noticeably
broadening of emission lines, it does not affect our ability to fit them
with Gaussian components.

\section{Results}\label{s:results}

Before analysing the individual spaxels/spectra we first examine
radially averaged, higher S/N measurements by combining spectra in
circular annuli. In each data cube we centre the annuli on the centre of
continuum flux distribution.  The width of the annuli is one spaxel
(0.67 arcsec) and their radii range from 1 to 19 spaxels in steps of 2
spaxels.  The emission lines in the radially averaged spectra are
analysed in exactly the same way as the individual spaxels described in
the previous section.

\subsection{\halpha / colour profiles}

\begin{figure*}
 \centerline{\psfig{figure=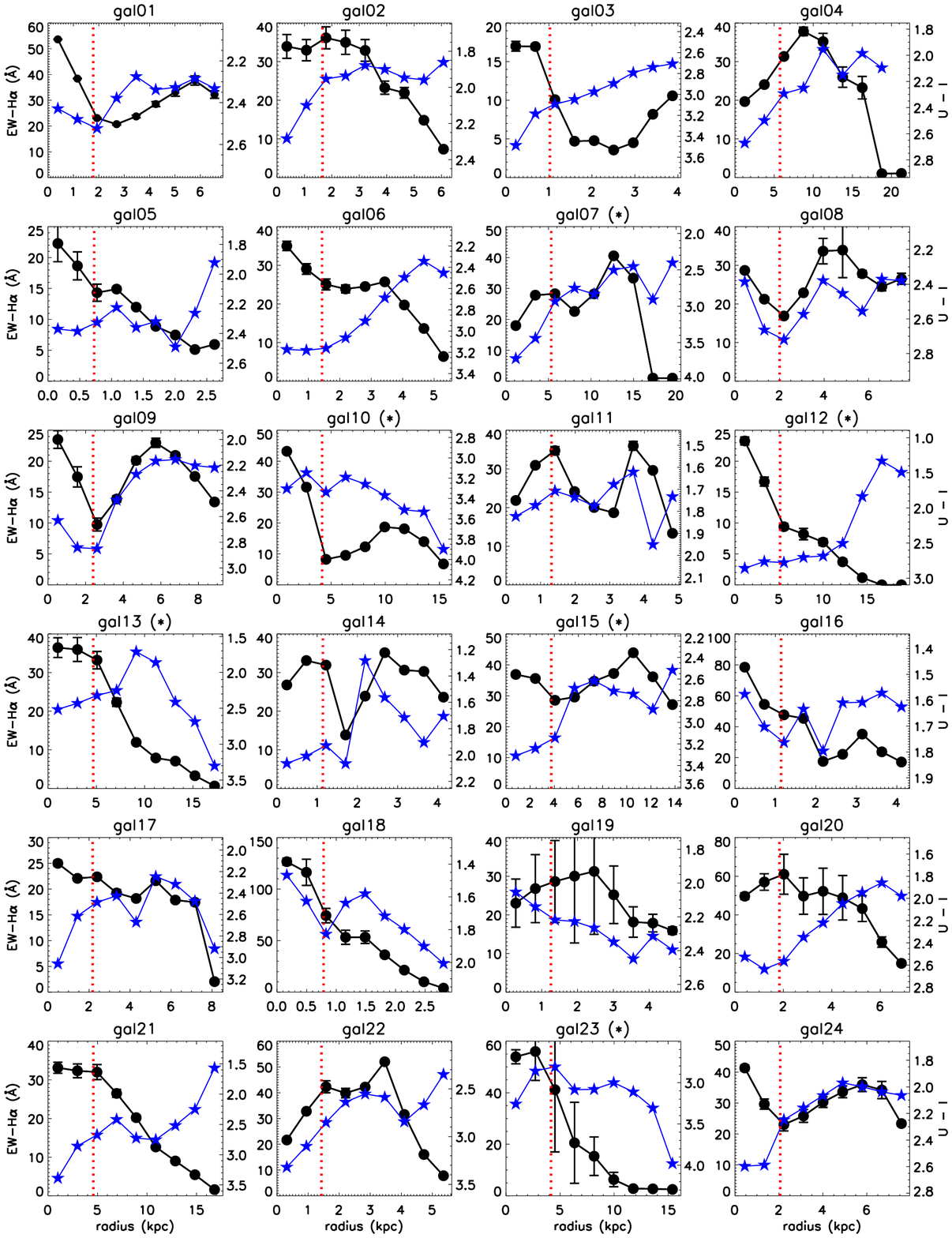,width=0.85\textwidth}}
 \caption{The radial profiles of EW[\halpha] for each of the galaxies in
   our sample (black lines and circles). 
   These quantities are derived in circular annuli of
   increasing radius. The dotted, red, vertical line shows the size of 
   a SDSS fibre. The large variety in profile shapes illustrates
   that \halpha\ emission and derived properties, such as star
   formation rate, are very sensitive to the choice of aperture. The
   error bars are propagated from errors in the fit.  The blue lines 
   and stars show the $u - i$ colour profiles derived from the SDSS images over
   the same annuli as the equivalent width profiles. The colour scales
   are shown on the right of each panel: note that this scale is 
   inverted to qualitatively match EW[\halpha]. Galaxies labelled with an asterisk
   are radio or BPT AGN, see text.}
\label{f:radprof}
\end{figure*}

Figure~\ref{f:radprof} shows the EW[\halpha] profiles derived by
simultaneously fitting three Gaussians to the \halpha + [NII] lines in
the summed spectra of each annulus. The errors are propagated from
errors in the fit.  \halpha\ emission is detected out to a radius
ranging between 3 and 20~kpc. For comparison, the SDSS fibre size
corresponds to 0.35--2.8~kpc depending on the redshift of the
galaxies. Thus, we trace on average the emission line flux out to a
radius $\sim$10 times that of the SDSS fibre size.

As the profile shapes are often slowly declining with radius or even
roughly constant, most of the \halpha\ flux in these systems is missed
by the SDSS spectra. For a qualitative comparison, u-i colour
profiles derived from the SDSS images and binned in the same annuli as
the emission, are also shown in this figure. These will be discussed
further in section~\ref{s:apeffects}.

\subsection{BPT}\label{sec:bpt}

Line ratio diagnostics are commonly employed to distinguish galaxies
with ionized gas consistent with HII regions from those requiring a
harder ionization field likely associated with the narrow line region of
an active galactic nuclei (AGN), although some of the parameter space
might also be consistent with radiation associated with shocks and
starbursts (Kewley et al, 2001, Sarzi et al, 2010).  Kauffmann et
al. 2003 (K03 hereon) and B04 have applied such diagnostics to the SDSS
spectra, applying semi-empirical divisions in the $\frac{\rm
  [OIII]\lambda5007}{H\beta}$ versus $\frac{\rm
  [NII]\lambda6584}{H\alpha}$ ``BPT'' diagram (Baldwin, Phillips \&
Terlevich, 1981) to classify galaxies into star-forming (left of the
empirical line marking the tight locus occupied by most galaxies), AGN
including LINERS (right of the Kewley et al. (2001) line calibrated to
maximal models of starburst galaxies) and composite (in between the two
lines) types.

\begin{figure*}
 \centerline{\psfig{figure=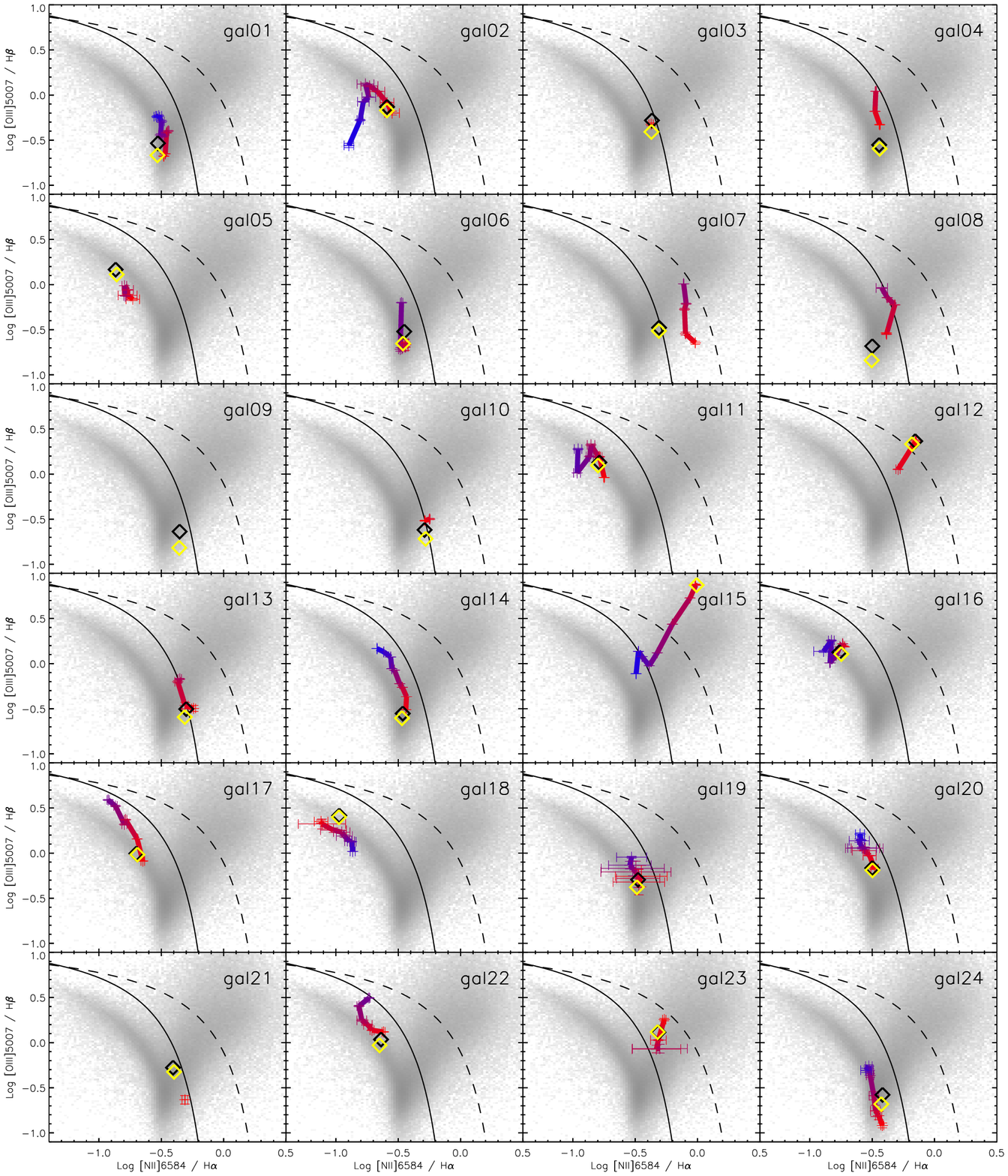,width=0.85\textwidth}}
 \caption{ The emission line flux ratios measured in the same annuli as
   in Fig~\ref{f:radprof} are shown here as 'tracks' on top of
   SDSS-derived BPT diagrams. The latter are taken from single-aperture
   measurements of some 500,000 galaxies in the SDSS database (B04).
   The radius increases by 2 spaxels for successive points along the
   track.  The direction of increasing radius is along the red to blue
   colour gradient for each track.  
   Although not selected as such, several systems have (hidden) AGN
   activity.  This is particularly striking in gal15 (Gerssen et
   al. 2009).  To verify our calibration we have also derived the line
   ratios directly from the spectra in the SDSS database. The results
   are shown as the yellow diamonds.  They are generally located close
   to the start of each track as expected. The black diamonds depict the
   line ratios obtained using the B04 line flux measurements. The dashed
   and solid lines show the boundary between star forming and AGN
   systems as derived by Kewley et al. (2001) and Kauffmann et
   al. (2003) respectively.}
 \label{f:bpt}
\end{figure*}

For radial bins where all four emission lines can be fit, we have
measured these ratios in our data.  In Fig.~\ref{f:bpt} the resulting
line ratio tracks are overplotted in colour on a greyscale background
showing the distribution of fibre-based SDSS measurements presented by
K03 and B04.  The empirical lines used by K03 and Kewley et al. (2001)
to divide pure star forming galaxies from AGN and composite galaxies are
overplotted in black.  The direction of increasing radius traced by the
line ratio tracks is from red to blue, with crosses marking measurements
for the individual annuli (radially binned with an annulus width of 2
pixels, corresponding to 1.5\arcsec).  To verify our calibration we have
also derived the line ratios directly from the spectra in the SDSS
database. These are shown as yellow diamonds. They are generally located
close to the red, low radius part of each track as expected, since they
cover only the central 3\arcsec.  As an additional check we also plot
the ratios obtained using the line fluxes derived by B04. These are
shown as black diamonds, and are located very close to the yellow
diamonds with small differences possibly due to differences in the
correction for underlying stellar balmer line absorption.

It is clear that our sample includes both galaxies for which the
ionization state is consistent with that expected in HII regions
(presumably normal star-forming galaxies) and those for which line
ratios require a component of harder ionizing radiation.  As noted in
Gerssen et al. 2009, gal15 contains an extended narrow line region
(ENLR), requiring a harder ionizing radiation component impacting the
ionized gas out to large $\sim9$kpc radii. Indeed, {\it all galaxies
  requiring harder ionization (7, 12, 15, 23) remain in that part of the
  diagram out to relatively large radii!}  Our targets were not in any
way selected to contain ENLRs, or even AGN. Therefore, at least based
upon our small, high EW[\Halpha] sample, ENLRs appear to be a common
phenomenon such that emission lines at large radii (i.e. $\sim1-10$ kpc)
are not independent of nuclear properties and as such do not trace only
star formation.

Ionizing radiation in galaxies whose tracks remain to the left of the
K03 dividing line should be dominated by star formation in
HII regions.  Consistent with this hypothesis, most of these galaxies
follow tracks along this star forming locus, verifying the consistency
of line ratios for resolved subsections of galaxies and radial bins
which extend far beyond the SDSS fibres. In galaxies
1,2,4,6,11,13,14,16,17,19,20,22 and 24, the data is consistent with a
movement to larger $\frac{\rm [OIII]\lambda5007}{H\beta}$ and smaller
$\frac{\rm [NII]\lambda6584}{H\alpha}$ (up and to the left) along the
locus with increasing radius.  This is consistent with expectations,
since in this direction metallicity is mostly decreasing and
ionization parameter increasing 
(see Fig.~4 of Levesque, Kewley \& Larson, 2010). 
Galaxies 2 and 5 move off the locus at large radii, where the signal
to noise becomes low in [NII] and [OIII] lines. Galaxy 18 presents a
reverse track, moving to smaller $\frac{\rm
  [OIII]\lambda5007}{H\beta}$ and larger $\frac{\rm
  [NII]\lambda6584}{H\alpha}$ at larger radii. This galaxy is unusual,
with a double nucleus probably related to a late merger stage.  During
a merger unusual gradients in metallicity and ionization parameter should
not be unexpected.

\subsection{Maps}
\label{s:maps}

In Fig.~\ref{f:maps1} we present maps derived from the emission line
analysis to the \halpha$+$[NII] lines in all individual spaxels/spectra.
We interactively inspected the best fit results in all spaxels and
flagged the poor fits.  The best-fit continuum and emission-line fluxes
are shown in columns one and two respectively.  The next two columns
show kinematical maps derived from the line fits, the velocity fields in
column three and the dispersion maps in column four.  The size of the
maps is the FoV of VIMOS on the sky, 27 by 27 arcsec, and is thus the
same in all panels and for all galaxies shown.  The horizontal black
bars in column one are five kpc long and are useful to gauge the
relative physical sizes of the galaxies in our sample (see also
Table~\ref{t:sample}).  In all maps we only plot those spaxels which are
not flagged and have Ha line flux ratio $\ge 3$.

\subsubsection{Continuum maps}

The colour SDSS images shown in Figure 1 have the same size and
orientation as the maps in Figure 9 and can be compared directly to the
continuum maps.  Galaxies gal13, gal14 and gal16 have a nearby
foreground star in their colour images that is also clearly noticeable
in their continuum maps.  The foreground stars in gal11 and gal17 are
not seen in the continuum maps as these regions were rejected in the
line-fitting analysis (too low S/N to fit emission lines).

\subsubsection{\Halpha\ line-flux maps}
\label{s:fluxmaps}

The galaxies in our sample were selected to be moderately gas-rich,
\Halpha\ EW $\ge 20$. Their spiral arm structure, apparent in Figure~1
and to a lesser extent in our continuum maps, can be readily traced in
the \Halpha\ line-flux maps.  In particular gal11 and gal24 display a
striking number of \Halpha\ clumps that partially trace the flocculent
spiral arms in these two systems.  Similarly, the grand design spirals
structure in galaxies gal06, gal07, gal10 and gal13 is also visible in
the line-flux maps.  And in systems that are close to edge-on the
\Halpha\ distribution traces spiral arm features where non can be seen
in broad band images.  The linear structure of \Halpha\ clumps seen in
gal17, for example, resembles an edge-on version of gal24.

Most off-center \Halpha\ emission is clumpy, and several systems in
particular, i.e. gal01, gal08, gal09, gal11, gal14, gal15, gal17, gal18,
gal22, contain clear off-center \Halpha\ clumps. We measure the flux of
off-centre clumps in these galaxies, fitting the \Halpha\ line for
summed spaxels -- the dominant error is the size of the clump.  These
are recorded along with estimated star formation rates using the
relation of Kennicutt 1998 (no dust correction) in
Table~\ref{t:clumps}. In the cases of gal11 and gal18 we record the flux
for all clumps, as there is no obvious nucleus.  These flux measurements
can be compared to the total and nuclear (SDSS aperture) fluxes
presented in table~\ref{t:flux} (see section~\ref{s:apeffects}). In all
cases the clumps are at least as luminous in \Halpha\ as the nucleus of
the galaxy. Thus, whilst most galaxies in our sample contain
\Halpha\ bright nuclei, this does not preclude the existence of brighter
star forming regions in the disk.  In fact, these clumps are bright
enough to drive a constant or increasing EW[\Halpha] with radius in
azimuthally averaged annuli, for many of the galaxies containing bright
off-center clumps, as shown in Fig~\ref{f:radprof}.

\begin{table}
\begin{center}
  \caption{Integrated off-centre clumps: \Halpha\ emission flux, and star
    formation estimated using the relation of Kennicutt 1998 (no dust
    correction).}
\vspace{0.1cm}
\begin{tabular}{ccc}
Name/clump & Flux & SFR \\
  & $10^{-17}erg s^{-1} cm^{-2}$ & $M_\odot yr^{-1}$ \\
\hline
gal01   & 755  & 0.120 \\
gal08   & 1171 & 0.241 \\
gal09   & 1052 & 0.318 \\
gal11SE & 290  & 0.025 \\
gal11W  & 258  & 0.022 \\
gal11N  & 344  & 0.029 \\
gal14   & 2149 & 0.022 \\
gal15   & 2503 & 2.040 \\
gal18E  & 2059 & 0.060 \\
gal18W  & 5710 & 0.167 \\
gal20   & 2459 & 0.419 \\
gal22E  & 523  & 0.053 \\
gal22W  & 510  & 0.052 \\
\end{tabular}
\label{t:clumps}
\end{center}
\end{table}

\subsubsection{\Halpha\ velocity fields}

The best-fit gas line-of-sight velocities are shown in column three.  In
general the \Halpha\ velocity maps are consistent with well defined,
regular rotation.  Even in systems that appear close to face-on in the
SDSS images and continuum maps, e.g. gal06, ga08 and gal11, low
amplitude rotation is still discernible.  In most galaxies in our sample
the observed velocity amplitudes lie in the range of 100 to 200 km/s.
In section~\ref{s:mass} we quantify the velocity amplitudes and use them
to derive a dynamical estimate of the enclosed mass in these systems.

\subsubsection{\Halpha\ dispersion maps}

The velocity dispersion maps shown in column four are corrected for
instrumental broadening by re-deriving our combined data cubes
\textit{without} sky subtraction, and then fitting a Gaussian function
to the 5577\AA\ sky line in order to measure the instrumental dispersion
for each spaxel.  This value is typically consistent with the measured
H$\alpha$ dispersion in the outer regions of galaxies, indicating that
these spaxels are dominated by instrumental dispersion.  We subtract
this from the measured H$\alpha$ width in quadrature, setting to 0 in
cases where the measured dispersion is lower than the instrumental
dispersion. Some galaxies show significant peaks in dispersion, centred
on the galaxy nucleus. We discuss this further in section~\ref{s:agn}.

\subsection{Enclosed mass} 
\label{s:mass}

The velocity fields shown in column~3 of Fig.~\ref{f:maps1} can be
used to obtain dynamical constraints on the enclosed mass in these
galaxies.  Building sophisticated kinematical models is outside the
scope of this paper.  Instead we obtain enclosed mass estimates by
fitting the velocity fields with a simple, infinitely thin circular
disk model.  The radial velocity distribution in this toy model is
characterized by a velocity that rises linearly from 0 km/s at the
center to a constant value, $V_{\rm circ}$, at a break radius, $R_{\rm
  break}$ and is constant thereafter.  The centre, position angle and
inclination of the velocity field model are also free parameters.  The
best fit model is found simply by varying the five model parameters
over a five-dimensional grid.  For each grid point we compute the
summed squared residual value of the difference between the model and
the observed velocity field.  As the best fit model we take the grid
point with the minimum residual value.

Not all \Halpha\ velocity fields in our sample are sufficiently well
defined, or sufficiently well sampled for our simple toy model to
produce an acceptable fit.  The aim of this toy model is to estimate a
dynamical mass using the best-fit circular velocity at a reference
radius of 10 kpc.  As we do not attempt to accurately constrain the
central behaviour of the velocities the exact value of the break
radius is not relevant here as long as $R_{\rm break} < 10$ kpc.  For
those galaxies in our sample that meet this criterion we list the
estimated dynamical mass in table~\ref{t:mass}.  

As a comparison, we list the stellar masses derived for K03 where these 
are available. They derive stellar mass to light ratios using the 
D4000\AA\ and H$\delta$ absorption features within the SDSS fibre, 
and apply this to the total z-band luminosity. 
To make a fair comparison, we compute an aperture correction by 
measuring the fraction of z-band flux within a central 10kpc aperture 
placed on the SDSS z-band images, masking nearby objects. The final 
column in table~\ref{t:mass} lists the stellar masses within 10kpc 
by applying this correction factor. 

\begin{figure*}
 \centerline{\psfig{figure=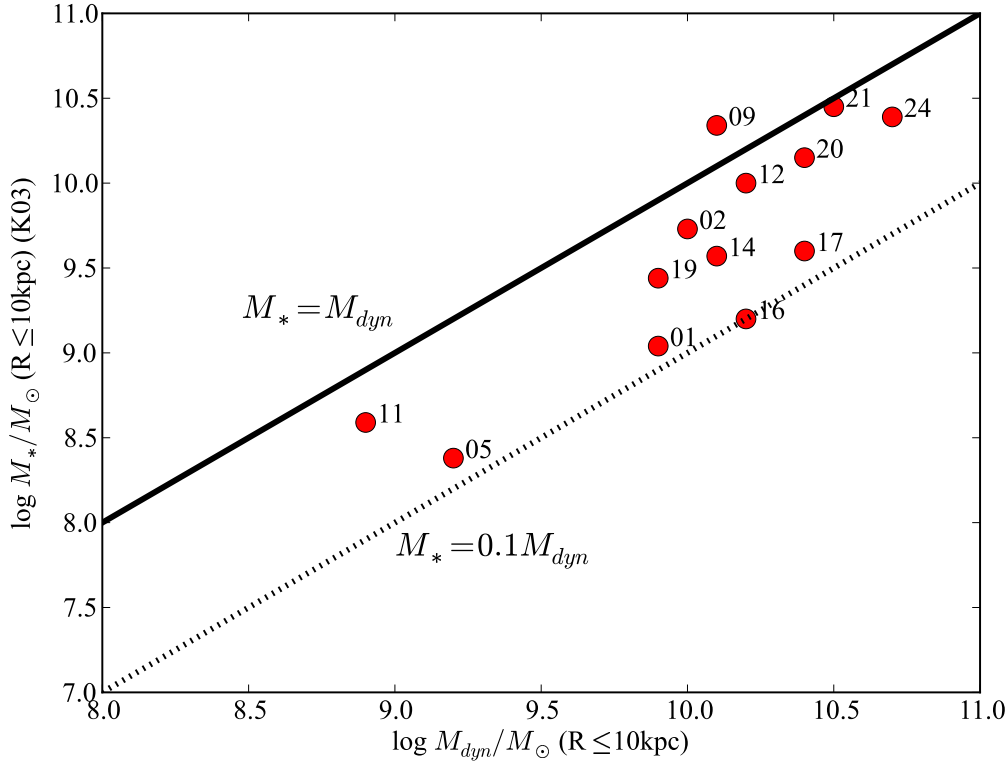,width=0.85\textwidth}}
 \caption{Dynamical mass from our toy model within 10kpc versus stellar
   mass from K03, scaled by z-band flux to that within a 10kpc circular
   aperture. Only galaxies for which these two parameters could be
   derived are included. The solid (dotted) line shows where the stellar
   mass accounts for all (one-tenth) the total mass within 10kpc. Our
   galaxies typically lie between these two lines, and this ratio shows
   no mass dependence.}
\label{f:masscomp}
\end{figure*}

Figure~\ref{f:masscomp} shows the dynamical versus stellar masses 
within an aperture of 10kpc. Solid and dotted lines respectively 
illustrate the locus of objects with 100\% and 10\% of the total 
mass in stars. Acknowledging uncertainties in both measurements, 
all galaxies are consistent with values in between these two extremes. 
Interestingly, there is no significant dependence of the fraction of 
mass in stars, on the mass itself for these measurements.

\begin{table}
\centering
\caption{Column 2 lists the enclosed total mass within 10kpc
  for those galaxies whose
  \halpha\ velocity fields can be fit with a simple model, see text.
  The third column lists the total stellar masses derived by K03. The
  fourth column corrects this to an aperture of 10kpc by computing the
  fraction of z-band light within this radius.}
\begin{tabular}{@{}lccc@{}}
 \hline
 Name & log $\frac{M_{dyn}}{M_\odot}(R\leq10 kpc)$ & \multicolumn{2}{c}{log $\frac{M_*}{M_\odot}$ (K03)} \\
      &						   & total & $(R\leq10 kpc)$ \\
\hline
gal01 &   9.9 &  9.333 & 9.04  \\
gal02 &  10.0 &  9.872 & 9.73  \\
gal03 &  --   &   --   & --    \\
gal04 &  --   & 10.930 & 10.53 \\
gal05 &   9.2 &  8.569 & 8.38  \\
gal06 &   --  &  9.575 & 9.47  \\
gal07 &   --  & 10.874 & 10.27 \\
gal08 &   --  & 10.016 & 9.83  \\
gal09 &  10.1 & 10.565 & 10.34 \\
gal10 &   --  & 11.022 & --    \\
gal11 &  8.9  &  8.817 & 8.59  \\ 
gal12 &  10.2 & 10.661 & 10.00 \\
gal13 &   --  & 10.964 & 10.41 \\
gal14 &  10.1 &  9.614 & 9.57  \\
gal15 &  --   &   --   & --    \\ 
gal16 &  10.2 &  9.239 & 9.20  \\
gal17 &  10.4 & 10.041 & 9.60  \\
gal18 &   9.8 &   --   & --    \\
gal19 &   9.9 &  9.537 & 9.44  \\
gal20 &  10.4 & 10.337 & 10.15 \\
gal21 &  10.5 & 10.925 & 10.45 \\
gal22 &  10.3 &   --   & --    \\	
gal23 &   --  & 11.220 & 10.87 \\
gal24 &  10.7 & 10.585 & 10.39 \\
\hline
 \label{t:mass}
\end{tabular}
\end{table}

%%%%%%%%%%%%%%%%%%%%%%%%%%%%%%%%%%%%%%%%%%%%%%%%%%%%%%%%%%%%%%%%%%%%%%%%

\section{Discussion}\label{sec:discussion}

\subsection{Radial gradients and SDSS aperture effects}\label{s:apeffects}

Figure~\ref{f:radprof} illustrates the wide range of EW[\halpha] and 
colour profiles for galaxies in our sample. 
The (u-i) colour profiles of some objects (e.g. 08,18) trace their 
EW[\halpha] profiles closely, suggesting that both trace the 
same features via the ratio of young to old stars. 

However most cases are not so simple: u-band light traces stars much 
less massive and older than the very massive and young stars traced 
by \halpha\ recombination (effectively the instantaneous SFR). 
\halpha\ emission is also highly sensitive to dust extinction in HII 
regions and can be triggered through different sources of ionizing 
radiation (e.g. AGN), whilst the slightly older stars dominating the 
u-band light will have migrated out of HII regions and are therefore 
sensitive mostly to diffuse dust absorption. At regions of low surface 
brightness and low star formation density (e.g. at large radius), 
the stochastic nature of star formation might lead to a significant 
excess of u-band light over that seen in the instantaneous SFR via \halpha. 
These and other effects must lead to the mismatch in the colour and 
\halpha\ profiles of most galaxies in our sample. 

The VIMOS data allow us to directly quantify aperture corrections by
comparing the \halpha\ line flux in the SDSS fibre to the total line
flux of the galaxy.  The derived line fluxes are listed in
table~\ref{t:flux} and have been corrected for absorption as described
in section~\ref{s:contsub}.  We simulate the SDSS apertures on our VIMOS
data cubes using the procedure described in
section\ref{s:reduction}. The derived \halpha\ line flux in the
simulated SDSS apertures are listed in column~2.  For comparison we
apply the same line fitting technique to the SDSS spectra themselves,
see column~3. The close correspondence between the derived line fluxes
validates our data reduction process which used the relative {\it
  continuum} level of the SDSS spectrum to recalibrate our flux scale.
The line fluxes derived by B04 in column~4 provide an independent
measurement.

The total line fluxes listed in column~5 are obtained by summing the 
flux from the radial binned spectra. The ratio of the line flux in
the simulated SDSS aperture to total flux is tabulated in column~6.
While the total flux is a measure within the field of VIMOS, the
\halpha\ maps in Fig.~\ref{f:maps1} demonstrate that this encompasses 
the brighter part of the disk, and any additional contribution to 
the total \halpha\ line flux should be negligible 
(By design as our selection criteria were constructed with that 
purpose in mind).

The large variation in this ratio demonstrates that generic aperture
corrections are of limited value when working with individual
galaxies. For instance, the ratios show no correlation with the factors
derived by B04 to correct aperture SFRs.  These were derived by assuming
a simple mapping of galaxy colour to emission line derived SFR
(calibrated using SDSS spectra with the inherent aperture effects).
This mapping is clearly unreliable for individual galaxies, as suggested
by Figure~\ref{f:radprof}. In many cases the central colour profile is
especially deviant from the EW[\halpha] profile - for example due to AGN
contamination or differential dust extinction.

\begin{figure*}
 \centerline{\psfig{figure=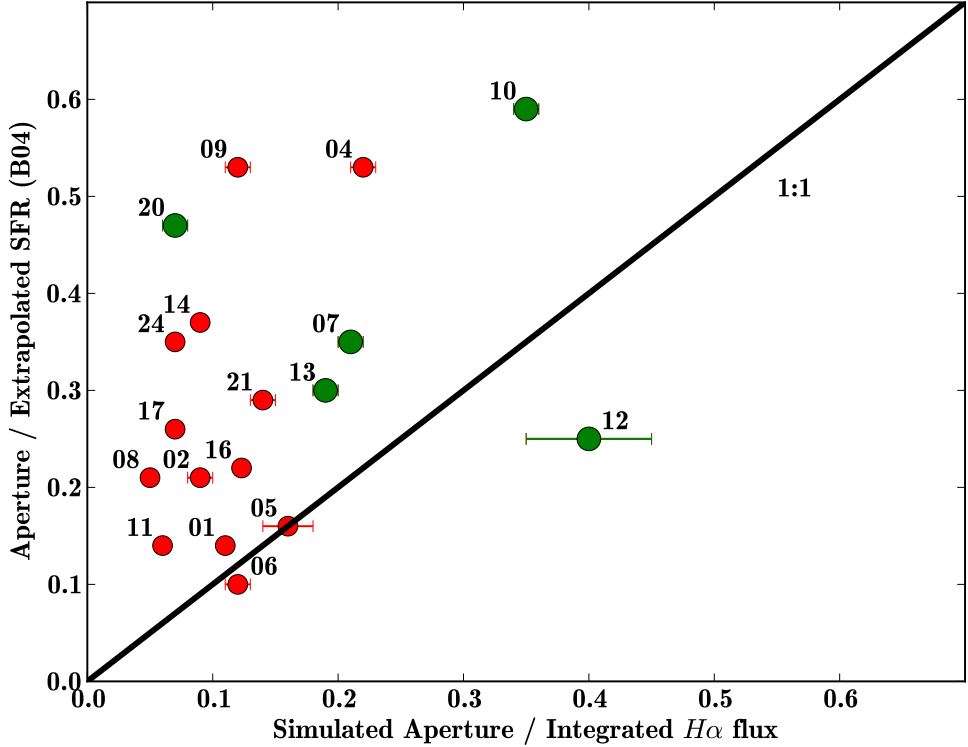,width=0.85\textwidth}}
 \caption{The measured ratio of \halpha\ flux measured inside the radius
   of a SDSS fibre (simulated SDSS fibre) to the total flux including
   errors (x-axis) plotted against the estimated SFR aperture correction
   from B04, (y-axis).  The green points are the AGN (radio or BPT). The
   solid black line shows the 1:1 relation. Galaxies 04, 15 and 18 have
   no SFR measurement from B04.  }
\label{f:apeffects}
\end{figure*}

Figure~\ref{f:apeffects} shows the aperture correction for \halpha\ line
flux calculated for our data plotted against the SFR aperture correction
of B04\footnote{Calculated from the fibre and total values
listed in files sfr\_dr4\_v2\_fib.fit and sfr\_dr4\_v2\_tot.fit respectively
obtained at: http://www.mpa-garching.mpg.de/SDSS/DR4/Data/sfr\_catalogue.html} 
Even ignoring the likely AGN (radio or BPT, shown in green - see
section~\ref{s:agn}), there is no correlation between the two
corrections, indicating that these colour-based corrections are
inaccurate for individual galaxies.  However we note that our sample
covers only galaxies with EW[\halpha]$_{fibre}>20$\AA\ and the
correction to \halpha\ flux need not equate exactly to the `SFR' correction.

As all points, except one AGN, fall to the left of the 1:1 relation in
Figure~\ref{f:apeffects}, it appears that B04 underestimate the
correction on average.  To quantify this we have examined the
distribution of y-axis value divided by x-axis value (the ratio of the
aperture fractions) for non-AGN only. We find that this distribution has
a mean at ratio~2.5, and scatter of 175\%. Thus the correction applied
by B04 appears -- on average -- underestimated, assuming that the dust
correction is not larger in the outskirts of star forming galaxies than
in the inner parts. To an order-of-magnitude level, the B04 aperture
correction may be valid.  However, for {\it individual} galaxies, we can
say that the systematic error induced by such a correction is a factor
2.5 for star forming galaxies, with even larger random errors,
especially where other sources of ionizing radiation are present.

\begin{table*}
\centering
\begin{minipage}{140mm}
\caption{\halpha\ line flux in units of $10^{-17}$ erg s$^{-1}$
  cm$^{-2}$.  VIMOS has a known spaxel near its centre with a low
  transmission.  This has affected our flux calibration in the central
  few spaxels of galaxies 03, 19 and 23 through the drizzling step in
  our reduction Hence, we do not attempt to compute the correction
  factors for these galaxies.}
\begin{tabular}{@{}lcccccc@{}}
 \hline
Name & Simul & SDSS & Brinchmann & Total & Simul/Total & Ap corr fac.\\
\hline
gal01 & 774$^{1)}$ (1.4)  & 203$^{2)}$ &  210 &  6905 (225) & 0.11 (0.003) & 0.14 \\
gal02 & 438 (44.9) & 490 &  542 &  4776 (332) & 0.09 (0.01)  & 0.21 \\
gal03 &            & 1045&  1101&  32621 (360)&              & 0.36 \\   
gal04 & 702 (7.5)  & 703 &  752 &  3166 (112) & 0.22 (0.01)  & 0.53  \\
gal05 & 229 (23.6) & 223 &  244 &  1422 (21)  & 0.16 (0.02)  & 0.16 \\
gal06 & 546 (28.3) & 557 &  612 &  4550 (103) & 0.12 (0.01)  & 0.10 \\
gal07 & 377 (13.4) & 394 &  434 &  1772 (26)  & 0.21 (0.01)  & 0.35 \\
gal08 & 334 (1.5)  & 318 &  347 &  6268 (379) & 0.05 (0.003) & 0.21 \\
gal09 & 877 (58.6) & 925 &  953 &  7185 (182) & 0.12 (0.01)  & 0.53 \\
gal10 & 957 (10.4) & 1063&  1121&  2727 (54)  & 0.35 (0.01)  & 0.59 \\
gal11 & 139 (1.1)  & 165 &  178 &  2276 (49)  & 0.06  (0.001)& 0.14 \\
gal12 & 179 (7.2)  & 246 &  259 &   448 (75)  & 0.40 (0.05)  & 0.25 \\
gal13 & 591 (48.6) & 530 &  580 &  3137 (49)  & 0.19 (0.01)  & 0.30 \\
gal14 & 1467 (0.1) & 1409&  1520&  16544 (17) & 0.09 (0.001) & 0.37 \\
gal15 & 849  (0.1) & 1094&  --  &  9853 (12)  & 0.09 (0.001) & --   \\ 
gal16 & 613 (0.1)  & 648 &  698 &  5051 (140) & 0.12 (0.003) & 0.22 \\
gal17 & 247  (4.1) & 260 &  275 &  3632 (52)  & 0.07 (0.001) & 0.26 \\
gal18 & 1080 (135.3)& 1012&  1062& 10356 (422)& 0.10 (0.01)  & --   \\
gal19 &            & 667 &  746 &             &              & 0.27 \\ 
gal20 & 603 (55.1) & 660 &  708 &  8383 (1138)& 0.07 (0.01)  & 0.47 \\
gal21 & 287 (23.2) & 285 &  299 &  2043 (40)  & 0.14 (0.01)  & 0.29 \\
gal22 & 294  (6.7) & 253 &  273 &  4293 (60)  & 0.07 (0.001) & -- \\ 
gal23 &            & 2244&  2398&             &              & 0.32 \\ 
gal24 & 441 (18.1) & 455 &  490 &  6737 (382) & 0.07 (0.004) & 0.35 \\
\hline
\multicolumn{7}{l}{1) Simul fibre at SDSS position has flux of 248.} \\
\multicolumn{7}{l}{2) SDSS fibre is not on gal01 nucleus.} \\ 
 \label{t:flux}
\end{tabular}
\end{minipage}
\end{table*}

\subsection{AGN}\label{s:agn}

\begin{figure*}

\centerline{\psfig{figure=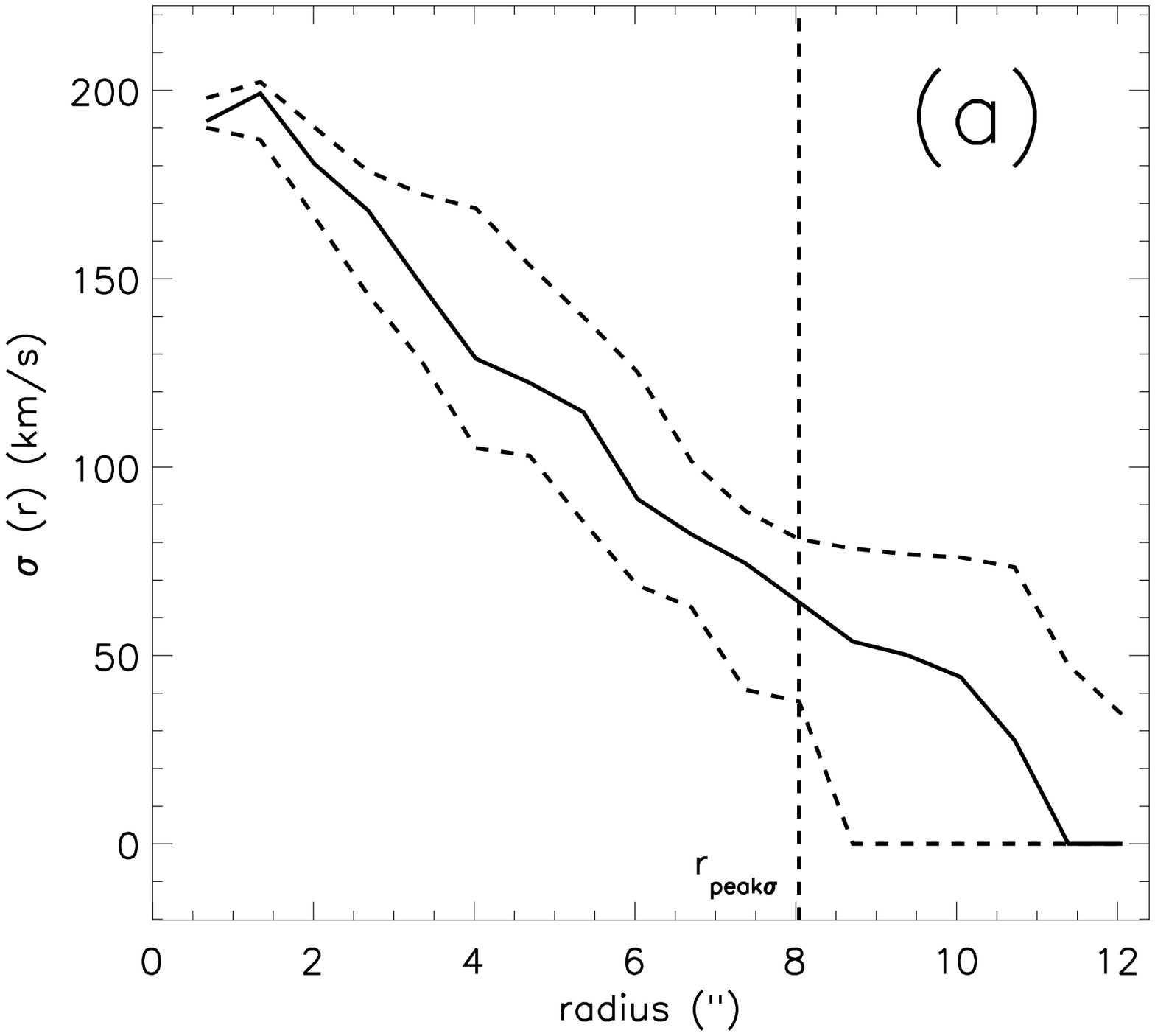,width=0.45\textwidth}\psfig{figure=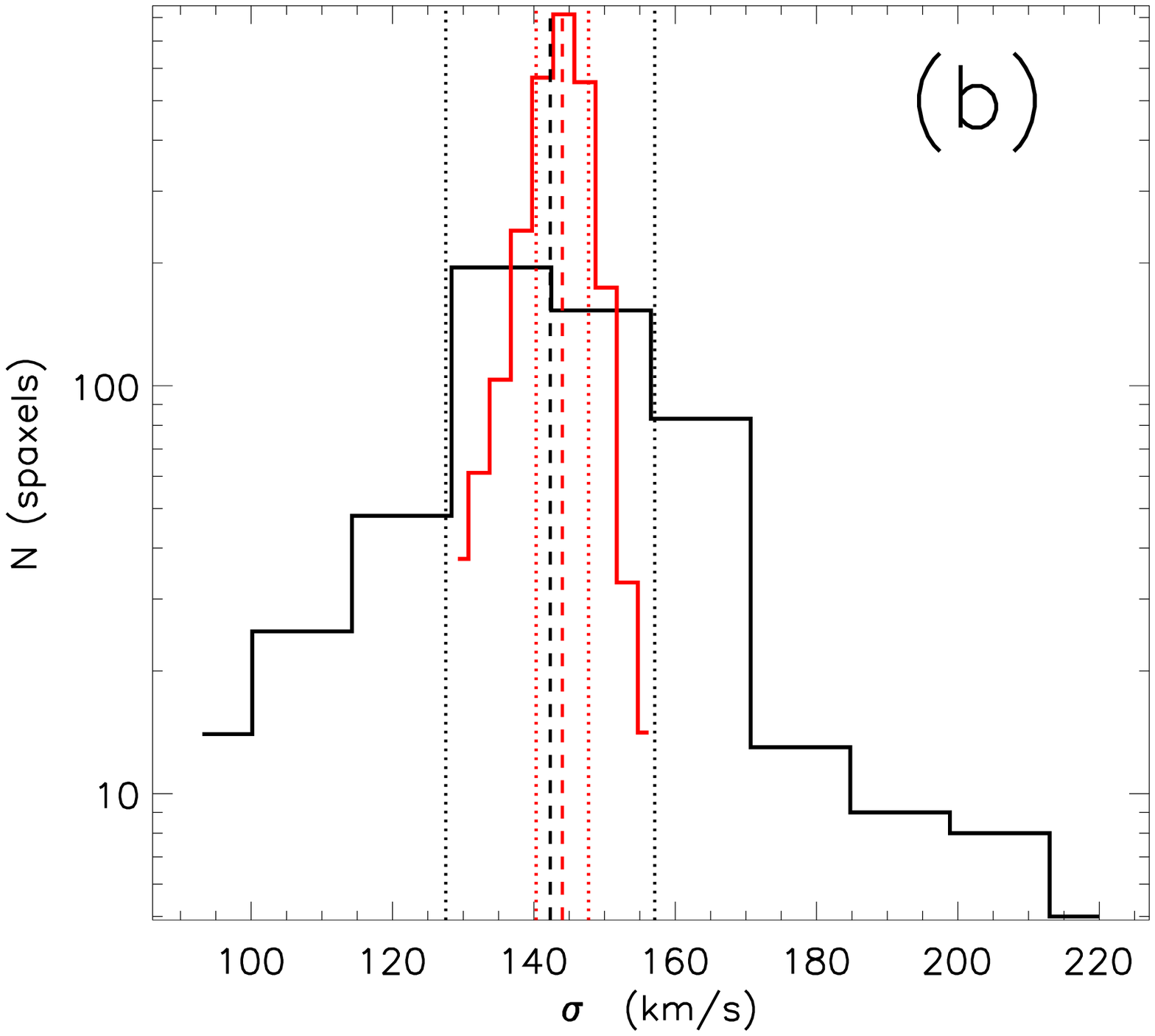,width=0.45\textwidth}}
\caption{Demonstration of our ``dispersion-peak'' diagnostics for the
  strongly dispersion-peaked galaxy gal15.  {\bf (a)} Dispersion
  profile, for spaxels binned by distance from galaxy centre, where
  the dispersion ($\sigma$) is the width of the \halpha\ line fit for
  each spaxel, in velocity units, and corrected for instrumental
  dispersion.  The solid and dashed lines represent the median and the
  25$^{th}$ and 75$^{th}$ percentiles of individual spaxel fits,
  respectively.  We define the radial extent of the central
  dispersion-peak, $r_{peak\sigma}$ (dashed vertical line), at the
  maximum radial bin within the 25$^{th}$ percentile of rises above
  zero ($\sigma \sim 0$ where the instrumental broadening dominates,
  and the uncertainty in $\sigma$ is larger than the intrinsic value).
  {\bf (b)} Histograms of $\sigma$ for spaxels at $r>r_{peak\sigma}$
  (in black) and of $\sigma_{\rm instr}$ -- the instrumental
  dispersion measured using a Gaussian fit to the 5577\AA\ sky line (in
  red).  The centroid of a Gaussian-fit to these histogram (dashed
  vertical lines) show that the H$\alpha$ line width for these spaxels
  is consistent with instrumental dispersion and the widths of the
  Gaussians (dotted lines at $\pm 1-\sigma$) show that the spaxel to
  spaxel scatter does not result from variation in $\sigma_{\rm
    instr}$ -- instead it probably results from errors in fitting the
  H$\alpha$ emission line, especially in regions of low flux.}
\label{f:sigprof15}
\end{figure*}

In section~\ref{sec:bpt} and figure~\ref{f:bpt} we saw that whilst
most galaxies in our sample have line ratios consistent with pure star
forming radiation, four clearly require a harder component of ionizing
radiation (7,12,15,23), whilst the line ratio tracks of galaxies 8, 10
and 13 are located above the locus of normal star forming objects,
close to the dividing line of (Kauffmann et al. 2003).
We will now examine other signatures of energy injection in the
central regions of our galaxies and assess the case for nuclear activity,
and its influence on the spatially resolved ionized gas and
other observables. Nuclear activity may be related to accretion onto
a super-massive black hole (an active galactic nucleus, AGN), although
certain configurations may also be explained by extreme star formation
(starbursts).

The National Radio Astronomy Observatory (NOAO) Very Large Array (VLA)
Faint Images of the Radio Sky at Twenty-Centimeters (FIRST: White et al, 97)
survey provides the best spatially resolved (1.8\arcsec\ pixels) radio map of the
sky (at the locations of our sources).  Radio fluxes are measured with a
typical rms of 0.15 mJy, with a 1 mJy source detection threshold.
Individual sources have 90\% confidence error circles of radius $<
0.5\arcsec$ at the 3 mJy level and 1\arcsec\ at the survey threshold.

Our sources have been matched to the FIRST radio catalogue using a
search radius equivalent to the VIMOS field of view.  The integrated
radio flux (Gaussian fit, White et al. 1997), and offset from the
position of the SDSS source, are tabulated in columns 2 and 3 of
table~\ref{t:agn}.  The nuclei of galaxies 10,13,15,20 and 23
contain significant radio sources which can clearly be seen in FIRST
images.  Small offsets (e.g. galaxy 15) are smaller than the radio
source size (despite formal positional errors below 1\arcsec) and
consistent with typical offsets in H$\alpha$ peak flux from the SDSS
position.  Radio sources with larger offsets (galaxies 3 and 4),
although within the VIMOS field of view, are unlikely to be associated
with the galaxy nuclei.

\subsubsection{Dispersion-Peaks}\label{s:peaks}

The maps illustrating the H$\alpha$ line width (last column of
figure~\ref{f:maps1}) demonstrate two regions of interest.  For the
majority of spaxels the measurement is limited by instrumental
resolution: the measured width of the 5577\AA\ sky-line is
$\sim3.3$\AA, which corresponds to $\sigma_{\rm instr} \sim 145 \rm km
s^{-1}$ for a typical galaxy at the wavelength of \halpha. To estimate
the intrinsic dispersion, the instrumental dispersion is subtracted in
quadrature from the measured dispersion -- in these regions the result
is close to zero and dominated by measurement errors.

However, in the central regions of many galaxies the \halpha\ line
width is broader, and resolved (a ``dispersion-peak'').  The thermal
dispersion of gas is $\sigma_{thermal} = \sqrt{\frac{k_B T}{\mu m_p}}$
which for $T\sim10^4$K results in $\sigma_{thermal} \sim 10~km s^{-1}$,
much lower than the measured dispersion. Projected circular velocity
variations within each spaxel can explain some enhancement, but not
much, especially for systems close to face-on. For gas to be so much
hotter requires an additional source of thermal or turbulent energy.
In central regions it seems likely that this can relate to the presence
of an AGN, although supernovae might also provide such energy.

We now compute several diagnostic parameters relating to the dispersion-peak,
presented in columns 4-7 in table~\ref{t:agn}.
To illustrate our method we provide figure~\ref{f:sigprof15},
demonstrating how these diagnostics are computed for a classic
dispersion-peak galaxy, gal15 (presented in Gerssen et al. 2009).

Individual spaxels are binned by radius
(distance of the spaxel from the centre, defined as in section~\ref{s:results}
to be the position of the continuum peak) in annuli of one pixel width.
Where the fit is acceptable, $\sigma$ is the measured width of the H$\alpha$ line for each spaxel,
converted to rest-frame velocity units.
For the spaxels contained within each radial bin,
we then examine the distribution of $\sigma$.

Figure~\ref{f:sigprof15}, panel {\bf(a)} shows the median (solid line),
and the 25$^{th}$ and 75$^{th}$ percentiles (dashed lines) of this distribution
versus radial bin. For galaxy 15 this $\sigma$-profile defines a
central peak, as seen in the $\sigma$-map of this galaxy (figure~\ref{f:maps1}).

In the outer regions of the galaxy instrumental broadening dominates
the measured line dispersion -- together with measurement error this
means the intrinsic dispersion is consistent with zero. This is seen in
the 25$^{th}$ percentile (lower dashed line) which goes to zero beyond
8\arcsec.  We use this fact to estimate the extent of the
dispersion-peak -- $r_{peak\sigma}$ is set to the maximum radial bin
before the 25$^{th}$ percentile goes to zero (effectively $\geq 25\%$
of the spaxels at larger radii have measured dispersion $\sigma \geq
\sigma_{\rm instr}$). $r_{peak\sigma}$ is marked with the vertical dashed
line in figure~\ref{f:sigprof15}, panel {\bf(a)}.

Figure~\ref{f:sigprof15}, panel {\bf(b)} shows the distribution of
measured dispersion ($\sigma_{\rm meas}$) for all spaxels containing
\halpha\ at $r>r_{peak\sigma}$ (black histogram). The red histogram
shows the distribution of $\sigma_{\rm instr}$ measured using the
5577\AA\ sky line for the same spaxels (using a smaller binsize, but
renormalized for comparison with the black histogram). The histograms
peak at roughly the same value of $\sigma_{\rm meas}$, which
demonstrates that the dispersion in these peaks is dominated by the
instrumental broadening.  To characterize these distributions we fit a
Gaussian to each peak.  The vertical dashed and dotted lines show the
median and $\pm 1-\sigma$ respectively, based on these fits. The
distribution of $\sigma_{\rm meas}$ is much broader than that for the
instrumental dispersion -- this shows that the scatter in $\sigma_{\rm
  meas}$ is dominated by fitting errors and not by spaxel to spaxel
variations in the instrumental dispersion.

Finally, we compute the overall significance of the dispersion-peak.
The significance that $\sigma_i$ is larger than then the baseline value
(computed using the mean dispersion for spaxels at $r>r_{peak\sigma}$ =
$<\sigma_{r \geq r_{peak\sigma}}>$) for each spaxel $i$ is:

\begin{equation}
 signif_i = \frac{\sigma_i - <\sigma_{r \geq r_{peak\sigma}}>}{rms(\sigma)}
\end{equation}

where the noise $rms(\sigma)$ is estimated by adding in quadrature the
scatter from the two distributions in measured and intrinsic sigma
(the black and red histograms in figure~\ref{f:sigprof15}, panel
{\bf(b)}) -- this is dominated by the measurement error from the
H$\alpha$ emission line fitting process ($\sigma_{\rm meas}$).

The combined significance of the peak is simply:

\begin{equation}
 signif(r < r_{peak\sigma}) = \sqrt( \Sigma_{i=1}^{N} signif_i^2 )
\end{equation}

for N spaxels with $r < r_{peak\sigma}$.

We also compute the intrinsic velocity dispersion $\sigma_{\rm cen}$ for
the central spaxel, by subtracting $\sigma_{\rm instr}$ from
$\sigma_{\rm meas}$ in quadrature.

A low amplitude ($\sigma_{\rm cen}$) but greatly extended
dispersion-peak such as galaxy 3 is highly significant. Such peaks may
result from the projected rotation curve convolved with the seeing.

It is physically more interesting to consider the physical origin of
higher dispersion peaks with significant S/N and size
($r_{peak\sigma}$).

Galaxies 10, 15 and 23 contain both significant dispersion-peaks with
$\sigma_{\rm cen}\geq 131$\kms and nuclear radio sources.  Of these,
galaxies 15 and 23 also have line ratios in the BPT diagram which
require harder ionization radiation, consistent with AGN origin.
Galaxies 7, 12 (and possible galaxy 10) have dispersion-peaks and
AGN-like line ratios, but no known radio counterpart. 

Most importantly, all galaxies with radio counterparts or clear AGN-like
line ratios have a central dispersion-peak, whilst all galaxies with no,
or very low significance dispersion-peaks have line ratios consistent
with a pure HII region ionization field and no radio source.  These
correlations strongly suggest that the same, or related physical
mechanism(s) are responsible for the energy injection which drive all
these phenomena (radio power, a hard ionization field and kinematic
heating).  The most promising candidate energy source is via coupling of
an AGN with the surrounding gas.  In this case radio emission traces
jets during the ``on'' phase of the jet duty cycle, whilst the
surrounding gas is ionized by the hard radiation of the accretion disk,
and can be kinematically heated by the jet
\citep[e.g.][]{vdb95,Koekemoer96}. And although the \Halpha\ flux of
off-centre clumps is frequently brighter than that in the nucleus, the
dispersion is never enhanced in those regions -- Fig~\ref{f:maps1}

Whatever the origin, these features are common in our sample (selected
only to have fibre EW[\halpha]$>20$\AA). Whilst no single selection method picks
up all objects, kinematic heating of the central ionized gas (dispersion-peaks)
selects a fairly complete subset of radio and BPT-selected sources, as well as
sources such as gal09 for which not all emission lines can be measured with a high
enough signal to noise ratio.

\begin{table*}
\centering
\begin{minipage}{140mm}
\caption{Properties of galaxies which might indicate the presence of
 an active galactic nucleus (AGN).  Columns 2 and 3: ``Deconvolved
 flux'' in mJy and offset of radio source from the SDSS optical
 position, where a galaxy has been matched to a radio source within
 the FIRST catalogue (White et al. 1997).  Note that FIRST fluxes are
 inexact but our main purpose is to illustrate the presence of radio
 flux.  Columns 4-7: The maps of H$\alpha$ emission line width
 ($\sigma$, see figures~\ref{f:maps1}-\ref{f:maps5} final column)
 have been examined for the presence of a ``dispersion-peak'' at the
 galaxy centre, corresponding to ionized gas which is heated to
 levels at which the H$\alpha$ line is resolved. Parameters describing
 this peak are presented here, and described in the text
 (section~\ref{s:agn}). Column 8: Are the line ratios consistent
 with an ionization field expected for pure HII regions? (Yes/No/?).}
\begin{tabular}{cccccccc}
\hline
Name & FIRST flux & FIRST offset & $\sigma_{\rm cen}$  &  $\sigma_{\rm instr}$ & $r_{\rm peak\sigma}$ &
 signif($r<r_{\rm peak\sigma}$) & BPT pure HII region?\\
& mJy & $\arcsec$ & km/s &  km/s & $\arcsec$ & & \\
\hline
gal01 & -    & -    &  69 & 148 & 3.4 &   8.7 & Y \\
gal02 & -    & -    & -   & 148 & -   & -     & Y \\
gal03 & 3.16 & 11.1 &  47 & 152 & 4.0 &  49.3 & Y \\
gal04 & 2.06 & 13.6 &  60 & 138 & 2.0 &   2.4 & Y \\
gal05 & -    & -    & -   & 138 & -   & -     & Y \\
gal06 & -    & -    &  25 & 138 & -   & -     & Y \\
gal07 & -    & -    & 173 & 135 & 3.4 &  30.4 & N \\
gal08 & -    & -    &  43 & 147 & 6.0 &  30.9 & ? \\
gal09 & -    & -    & 121 & 148 & 6.0 &  42.5 & Y \\
gal10 & 2.72 & 0.8  & 146 & 142 & 2.7 &  22.6 & ? \\
gal11 & -    & -    & -   & 142 & -   & -     & Y \\
gal12 & -    & -    &  84 & 139 & 1.3 &   5.3 & N \\
gal13 & 2.31 & 0.4  &  79 & 141 & 6.0 &  18.7 & ? \\
gal14 & -    & -    & -   & 141 & -   & -     & Y \\
gal15 & 3.03 & 2.5  & 192 & 144 & 8.0 &  81.3 & N \\
gal16 & -    & -    & -   & 144 & -   & -     & Y \\
gal17 & -    & -    &   8 & 148 & 0.7 &   0.4 & Y \\
gal18 & -    & -    &  29 & 148 & -   & -     & Y \\
gal19 & -    & -    & -   & 148 & -   & -     & Y \\
gal20 & 3.87 & 1.3  &  16 & 147 & 1.3 &   0.7 & Y \\
gal21 & -    & -    &  66 & 140 & 2.7 &   7.6 & Y \\
gal22 & -    & -    & -   & 140 & -   & -     & Y \\
gal23 & 6.22 & 0.7  & 131 & 141 & 4.7 &  24.9 & N \\
gal24 & -    & -    & -   & 147 & -   & -     & Y \\

\hline
\label{t:agn}
\end{tabular}
\end{minipage}
\end{table*}

\section{Summary}\label{s:summary}

In this data paper we have presented VIMOS IFU data of 24 galaxies
selected from the SDSS database We use our VIMOS maps to quantify the
differences in simulated SDSS apertures to integrated \Halpha\ flux.
These flux ratios are uncorrelated with aperture corrections based on
simple colour based SDSS extrapolations which are underestimated by a
factor $\sim$2.5 on average, with 175\% scatter with respect to measured
ratios.
We have estimated the enclosed dynamical mass from toy model fits to the
\Halpha\ velocity fields finding between 10 - 100\% of mass at $<10$ kpc
in stars.
We examine various signatures of nuclear activity (line ratios,
dispersion peaks, radio emission) and find that about half of the
galaxies in our sample show no evidence for nuclear activity or
non-thermal heating.
As our sample of star forming galaxies is spatially well sampled it
can be used as a local reference sample for a comparison with high-z
galaxies (e.g. the SINS survey, F\"orster-Schreiber et al. 2009).
Lessons should be learned at low redshift where we have the higher
spatial resolution, signal to noise, and ancillary data to attribute
cause and effect.  For instance, ionized gas kinematics are largely
driven by non-gravitational processes such as turbulence.
The next generation of observatories, including JWST, ALMA and ELTs,
will focus on observations of the high redshift Universe.  The
availability of detailed observations in the nearby Universe will be
of great value to facilitate the interpretation of high-z data.
High spatial resolution, wide-field IFS surveys of ionized gas providing
samples such as the one presented in this paper, are ideally suited to
the task of building up local reference samples. \\

%%%%%%%%%%%%%%%%%%%%%%%%%%%%%%%%%%%%%%%%%%%%%%%%%%%%%%%%%%%%%%%%%%%%%%%%

\begin{figure*}
 \centerline{\psfig{figure=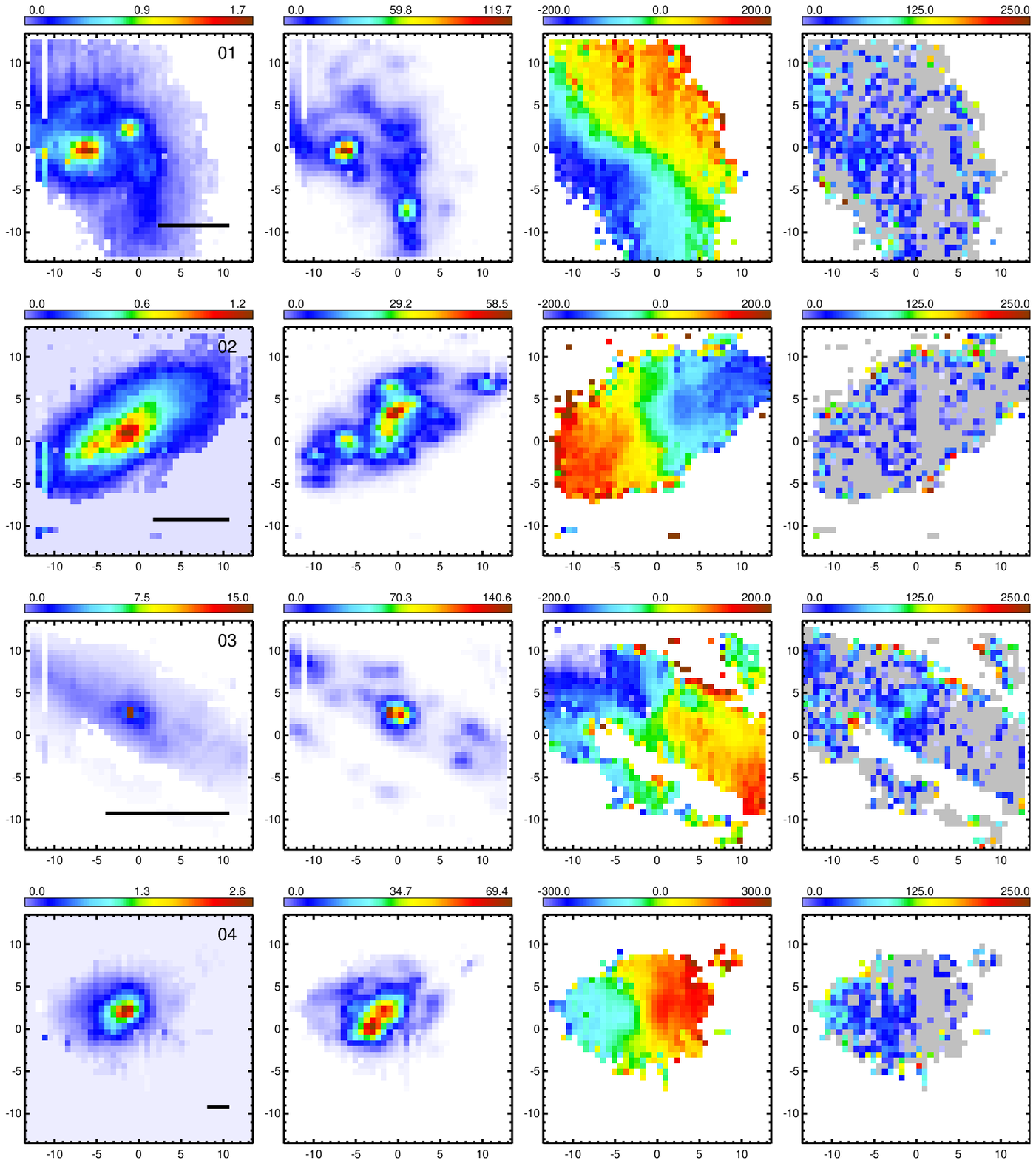,width=0.8\textwidth}}
 \vspace{0.5cm}
 \caption{Properties of the 24 SDSS galaxies in our VIMOS IFU sample.
   The galaxies are plotted in numerical order from top to bottom.  The
   results shown in these panels are derived by fitting the
   H$\alpha$+[NII] emission lines with a set of three Gaussians.  In all
   maps only those spaxels with S\/N ratios $\ge 3$ are plotted.  The
   best-fit continuum level maps are shown in {\bf column 1} and the
   H$\alpha$ line flux maps in {\bf column 2}. Flux is in units of
   $10^{-17}$ erg s$^{-1}$ cm$^{-2}$.  
   The velocity fields and velocity dispersion maps in km s$^{-1}$ are
   shown in {\bf column 3} and {\bf column 4} respectively.  Note that
   in the instrumental dispersion profile has been subtracted in
   quadrature from the best-fit dispersion measurements.  All colour
   maps are linear and the orientation in each panel is such that North
   is up and East to the left with the axis scales in arcsec.  Spaxels
   that are grey in column 4 have measured widths that are below the
   measured sky line widths. (Regions of low flux where fitting errors
   dominate measurements of dispersion.)
   The horizontal black bars in the first column are all five kpc long. }
 \label{f:maps1}
\end{figure*}

\addtocounter{figure}{-1} 
\begin{figure*}
 \centerline{\psfig{figure=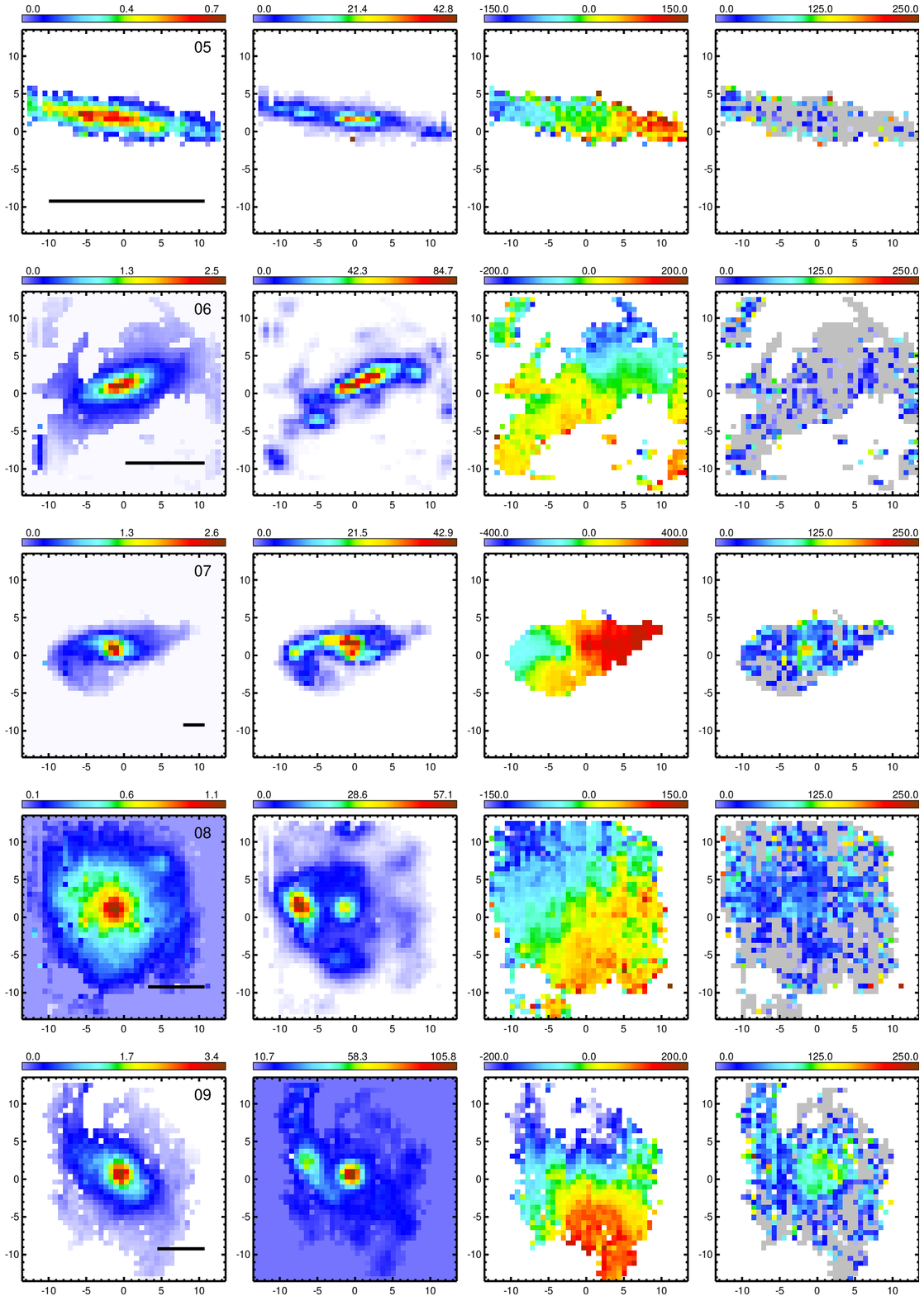,width=0.8\textwidth}}
 \caption{cont.
 \label{f:maps2}}
\end{figure*}

\addtocounter{figure}{-1} 
\begin{figure*}
 \centerline{\psfig{figure=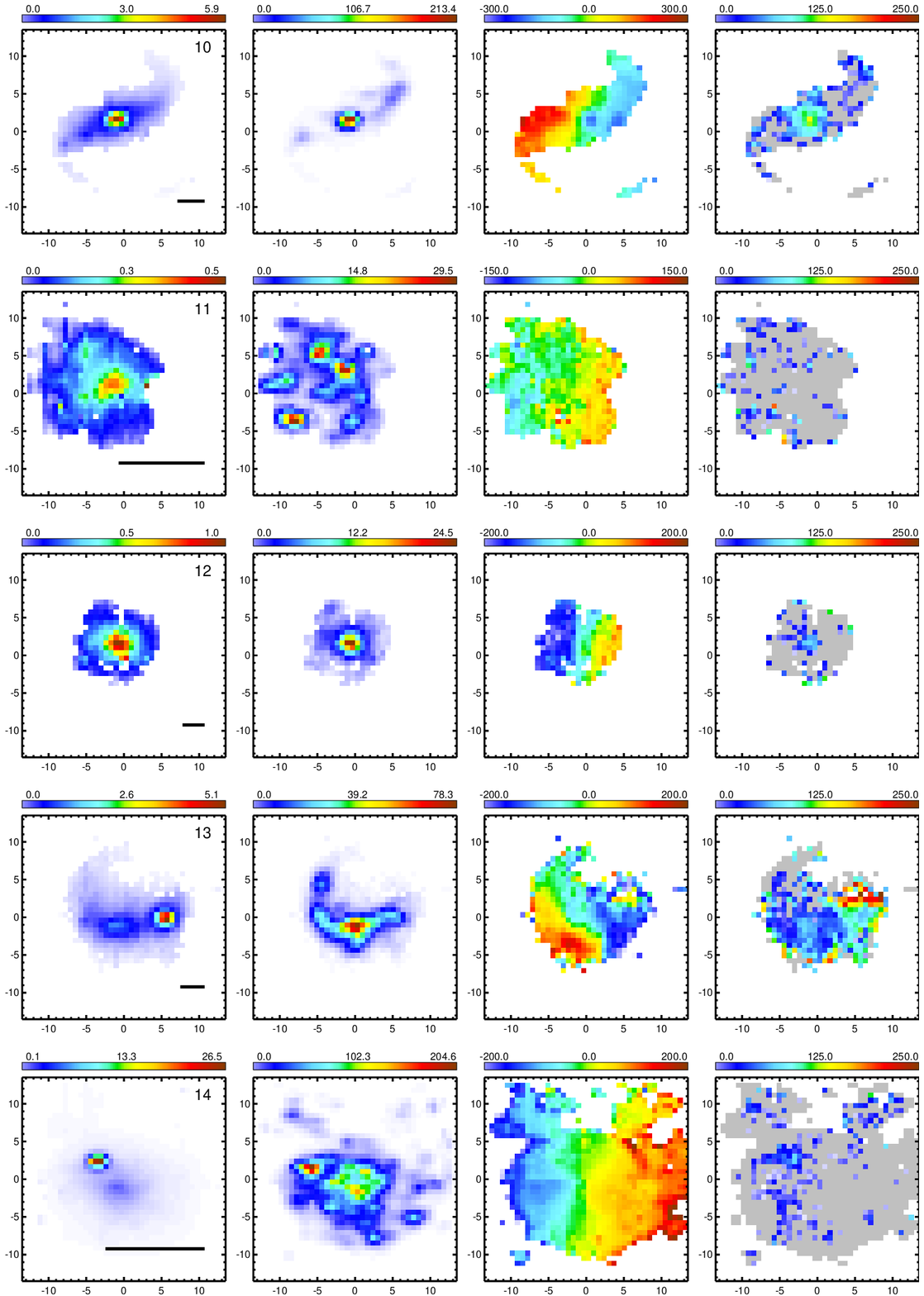,width=0.8\textwidth}}
 \caption{cont.
 \label{f:maps3}}
\end{figure*}

\addtocounter{figure}{-1} 
\begin{figure*}
 \centerline{\psfig{figure=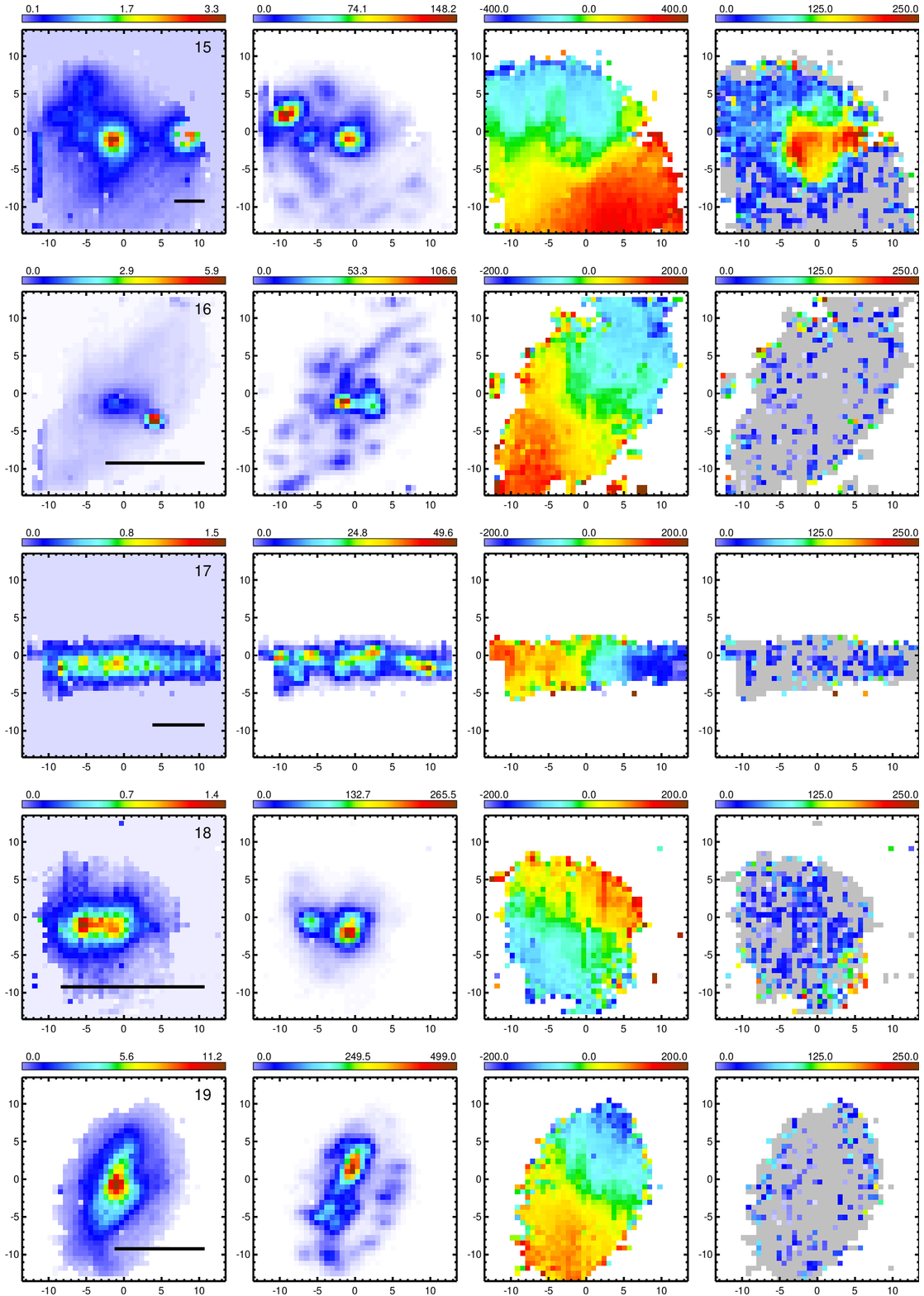,width=0.8\textwidth}}
 \caption{cont.
 \label{f:maps4}}
\end{figure*}

\addtocounter{figure}{-1} 
\begin{figure*}
 \centerline{\psfig{figure=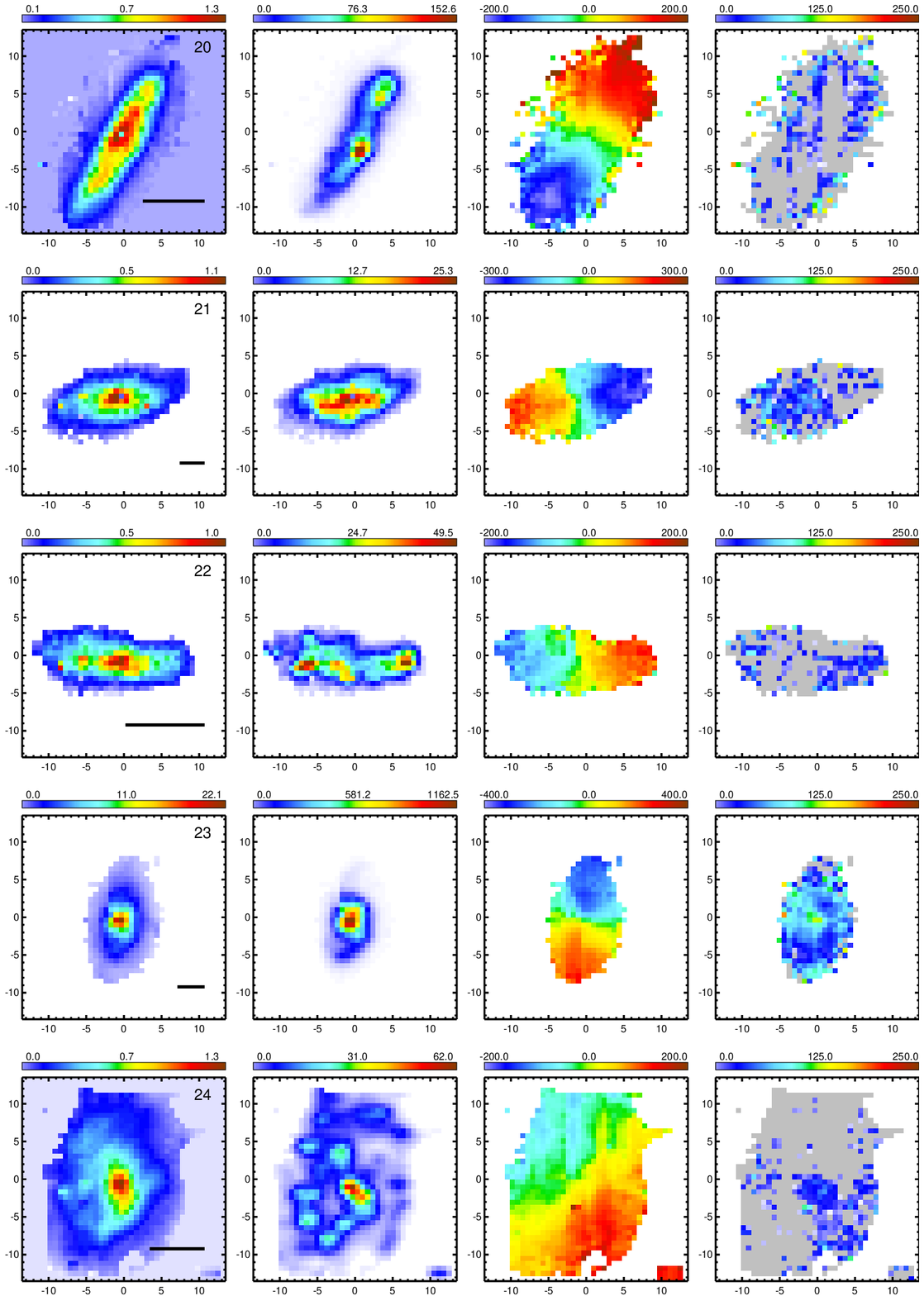,width=0.8\textwidth}}
 \caption{cont.
 \label{f:maps5}}
\end{figure*}

%%%%%%%%%%%%%%%%%%%%%%%%%%%%%%%%%%%%%%%%%%%%%%%%%%%%%%%%%%%%%%%%%%%%%%%%

\section*{Acknowledgements}

We thank Richard Bower for his contribution to the early stages of the
work presented here. Daniel Kupko has been instrumental in fitting the
continuum in our radially binned spectra.  We also like to thank Bodo
Ziegler and the referee for comments and suggestions that helped to
improve the paper.

Funding for the Sloan Digital Sky Survey (SDSS) and SDSS-II has been
provided by the Alfred P. Sloan Foundation, the Participating
Institutions, the National Science Foundation, the U.S. Department of
Energy, the National Aeronautics and Space Administration, the Japanese
Monbukagakusho, and the Max Planck Society, and the Higher Education
Funding Council for England. The SDSS Web site is http://www.sdss.org/.

The SDSS is managed by the Astrophysical Research Consortium (ARC) for
the Participating Institutions. The Participating Institutions are the
American Museum of Natural History, Astrophysical Institute Potsdam,
University of Basel, University of Cambridge, Case Western Reserve
University, The University of Chicago, Drexel University, Fermilab, the
Institute for Advanced Study, the Japan Participation Group, The Johns
Hopkins University, the Joint Institute for Nuclear Astrophysics, the
Kavli Institute for Particle Astrophysics and Cosmology, the Korean
Scientist Group, the Chinese Academy of Sciences (LAMOST), Los Alamos
National Laboratory, the Max-Planck-Institute for Astronomy (MPIA), the
Max-Planck-Institute for Astrophysics (MPA), New Mexico State
University, Ohio State University, University of Pittsburgh, University
of Portsmouth, Princeton University, the United States Naval
Observatory, and the University of Washington.

\label{lastpage}

\end{document}